\documentclass[english]{article}
\usepackage[utf8]{inputenc}
\usepackage{dsfont}
\usepackage{amsmath}
\usepackage{amsfonts}
\usepackage{amssymb}
\usepackage{amsthm}
\usepackage{authblk}
\usepackage{booktabs}
\usepackage[title]{appendix}
\usepackage[T1]{fontenc}
\usepackage{cite}
\usepackage{enumerate}
\usepackage{graphicx}
\usepackage[left=2cm,right=2cm,top=2cm,bottom=2cm]{geometry}
\usepackage{hyperref}
\usepackage{mathabx}
\usepackage{multirow}
\usepackage[normalem]{ulem}
\usepackage{scalerel}[2014/03/10]
\usepackage{stackengine}
\usepackage{subcaption}
\usepackage{titlesec}

\makeatletter
\newcommand\ackname{Acknowledgements}
\if@titlepage
   \newenvironment{acknowledgements}{%
       \titlepage
       \null\vfil
       \@beginparpenalty\@lowpenalty
       \begin{center}%
         \bfseries \ackname
         \@endparpenalty\@M
       \end{center}}%
      {\par\vfil\null\endtitlepage}
\else
   \newenvironment{acknowledgements}{%
       \if@twocolumn
         \section*{\abstractname}%
       \else
         \normalsize
         \begin{center}%
           {\bfseries \ackname\vspace{0mm}\vspace{\z@}}%
         \end{center}%
         \quotation
       \fi}
       {\if@twocolumn\else\endquotation\fi}
\fi
\makeatother

\usepackage{color}

\newcommand{\diff}{\mathrm{d}}
\newcommand{\ide}{1\hspace{-1mm}{\rm I}}
\newcommand\reallywidehat[1]{\ThisStyle{%
		\setbox0=\hbox{$\SavedStyle#1$}%
		\stackengine{-1.0\ht0+.5pt}{$\SavedStyle#1$}{%
			\stretchto{\scaleto{\SavedStyle\mkern.15mu\char'136}{2.6\wd0}}{1.4\ht0}%
		}{O}{c}{F}{T}{S}%
}}

\begin{document}

\title{Generalized Uncertainty Principle: from the harmonic oscillator to a QFT toy model
}

\author[1]{Pasquale Bosso\thanks{pasquale.bosso@uleth.ca}}
\author[2,3]{Giuseppe Gaetano Luciano\thanks{gluciano@sa.infn.it}}
\affil[1]{Theoretical Physics Group and Quantum Alberta,\protect\\ University of Lethbridge, 4401 University Drive, Lethbridge, Alberta, Canada, T1K 3M4\vspace{0.5em}}
\affil[2]{Dipartimento di Fisica, Universit\`a degli Studi di Salerno,\protect\\ Via Giovanni Paolo II, 132 I-84084 Fisciano (SA), Italy\vspace{0.5em}} 
\affil[3]{INFN, Sezione di Napoli, Gruppo collegato di Salerno,\protect\\ Via Giovanni Paolo II, 132 I-84084 Fisciano (SA), Italy.}

\date{\today}
\setcounter{equation}{0}
\maketitle

\begin{abstract}
Several models of quantum gravity predict the emergence of a minimal length at Planck scale. This is commonly taken into consideration by modifying the Heisenberg Uncertainty Principle into the Generalized Uncertainty Principle.
In this work, we study the implications of a polynomial Generalized Uncertainty Principle on the harmonic oscillator. We revisit both the analytic and algebraic methods, deriving the exact form of the generalized 
Heisenberg algebra in terms of the new position and momentum operators.
We show that the energy spectrum and eigenfunctions are affected in a non-trivial way.
Furthermore, a new set of ladder operators is derived which factorize the Hamiltonian exactly.
The above formalism is finally exploited to construct a quantum field theoretic toy model based on the Generalized Uncertainty Principle.

\end{abstract}

\section{Introduction}
Understanding how to merge quantum theory and general relativity is one of the most demanding challenges in modern theoretical physics. Despite the sustained  efforts~\cite{QG1,QG2,QG3,QG4,QG5,QG6,AdSan,QG8,QG9,QG10}, a consistent framework is still missing, the major obstacle  being the lack of experimental evidences at the scale where quantum gravity (QG) effects are expected to become manifest, \emph{i.e.} the Planck scale. Among the most robust predictions of QG models, the emergence of a minimum length of the order of Planck length  $\ell_p\sim10^{-35}$ m appears as  a distinctive signature of a
classical-to-quantum transition of gravity. This has found confirmation in a series of {\it gedanken} experiments involving the formation of gravitational
instabilities in high-energy string scatterings~\cite{QG1,QG2,QG3,QG5} and the creation of micro black holes~\cite{ScardBH}.  
Hence, in the would-be theory of QG, Planck length marks a threshold beyond which the common understanding of spacetime 
as a smooth continuum breaks down, giving way to a granular foamy structure due to inherent quantum fluctuations.

Implications of theories with a fundamental length are often 
addressed by deforming the Heisenberg Uncertainty Principle (HUP) in such a way as to accommodate a minimum uncertainty in position at Planck scale. The
ensuing relation, commonly known as Generalized Uncertainty Principle (GUP),
has been studied in a variety of contexts, ranging from quantum theory~\cite{KMM,DasPrl,Jizba,FrPan,HusQFT,BossoCohe,ScardLamb,ScardLuc,Lake:2018zeg,LucPetr,BossoDas:2020,Blasone:2019wad,Shababi,ShabaLuc,Luciano:2021cna,Bosso:2020aqm}, to black hole physics~\cite{BH1,BH2,BH3,BH4,BH5,BH5bis,BH6,BH6bis,BH7,BH8,BH9,BH10,BH11,BH12} and cosmology~\cite{GUPcosm1,GUPcosm2}. Potential effects of a modified Heisenberg algebra have also been explored in graphene~\cite{Graphene}, where the minimal length
is provided by the honeycomb lattice spacing. From theoretical shores, in recent years the investigation of the GUP has landed on more phenomenological grounds. In this vein, tests of GUP physics have been proposed in quantum optics~\cite{Bruk}, quantum mechanics~\cite{AliTest,Bawaj,Pendu} and gravity scenarios~\cite{GravBar,ScarCas, BossoLigo}, among others. 

Due to the relatively simple structure and the wide range of physical applicability,
a paradigmatic system where to study GUP effects is the harmonic oscillator (HO). The analytic resolution of the linear HO in the context of the GUP has been first considered in~\cite{KMM}, deducing both the modified energy spectrum and the corresponding eigenfunctions. On the other hand, in~\cite{Pedram:2012ui} 
the algebraic method has been revisited to construct the exact HO coherent states, the weight functions and the related probability distributions.  This has been achieved by resorting to the modified Heisenberg algebra of Ref.~\cite{Curado} and exploiting only the formal action of the generalized lowering and raising operators on the HO Fock space states. In passing, we mention that a perturbative construction of the generalized HO ladder operators has been exhibited in~\cite{Bosso:2018syo} for the case of an arbitrary self-adjoint polynomial perturbation of the commutator. Moreover, various forms of deformed oscillator algebra have been investigated in~\cite{Oh,Quesne} in the context of $q$-Hopf structure. 

Starting from the above premises, in this work we study the HO within the framework of a polynomial GUP. First, we solve  the stationary Schr\"odinger equation analytically and evaluate GUP corrections to the HO eigenvalues and eigenfunctions.
This paves the way for an \emph{ab initio} construction of the generalized HO Heisenberg algebra, which leads to the exact derivation of the ladder operators as functions of the GUP position and momentum.
We wish to emphasize that the novelty of our approach compared to Refs.~\cite{Curado,Pedram:2012ui} is that it allows featuring HO eigenstates not only in terms of their number of quanta and energy content, but also in terms of the expectation
value of the  position and momentum in the presence of a minimal length. This is especially useful in the study of Planck scale effects on high-precision oscillator frequency measurements and, in particular, in the computation of the trajectory of the oscillator in a (generalized) coherent state, as recently discussed in Ref.~\cite{Plenio}. Besides its own interest, we show that the above formalism can be exploited to construct a quantum-field-theoretic toy model based on the GUP. To avoid technicalities, we focus on the simplest case of a real free scalar field in $1+1$ dimensions. The extension to higher dimensions, as well as to the case of the Dirac field, deserves careful attention and will be investigated elsewhere.  

The layout of the paper is as follows: in Sec.~\ref{Mathframe}, we set the mathematical framework of the GUP. In Sec.~\ref{Analytic}, we solve the GUP harmonic oscillator by using both the analytic and algebraic methods.  Inspired by these results, in Sec.~\ref{QFT} we propose a field-theoretic toy model to quantize the Klein-Gordon field with the GUP. Conclusions and outlook are discussed in Sec.~\ref{Conc}. Two  Appendices contain computational details. Throughout the manuscript we set $c=1$, while we keep $\hbar$ explicit.

\section{GUP: the mathematical framework}
\label{Mathframe}
In this Section we introduce some basic mathematical tools that are necessary  for our next analysis of the GUP framework.
We start by assuming that the commutator between two given physical quantities $\hat q$ and $\hat p$  can be expressed in terms of a function of one of them as
\begin{equation}
	[\hat{q},\hat{p}] = i \hbar f(\hat{p}). \label{eqn:GUP}
\end{equation}
With a view to identifying $\hat q$, $\hat p$ with the physical position and momentum operators of our system and Eq.~\eqref{eqn:GUP} with the modified commutator 
modeling gravity corrections in the Planckian regime, we assume that $f(\hat p) \to \ide$ when the characteristic momentum is much smaller than the Planck momentum, 
so as to reproduce the standard Heisenberg relation at low-energy scale.

Clearly, $\hat q$ and $\hat p$ are not conjugated variables.
Let us then say that there exists a pair of variables $\hat Q$ and $\hat P$ such that
\begin{equation}
	[\hat{Q},\hat{P}] = i \hbar.
\end{equation}
We want to find $\hat{q}$ and $\hat{p}$ in terms of $\hat{Q}$ and $\hat{P}$.
Thus, by considering $\hat{q} = q(\hat{Q},\hat{P})$ and $\hat{p} = p(\hat{Q},\hat{P})$, we have (see Appendix~\ref{funcanvar})
\begin{equation}
	[q(\hat{Q},\hat{P}),p(\hat{Q},\hat{P})] = \frac{1}{2} \sum_{\substack{j,j'=0\\j+j'>0}}^\infty \frac{(i \hbar)^j (- i \hbar)^{j'}}{j! j'!} \left[\widehat{\frac{\partial^{j+j'} p}{\partial Q^{j'} \partial P^j}} \widehat{\frac{\partial^{j+j'} q}{\partial Q^j \partial P^{j'}}} - \widehat{\frac{\partial^{j+j'} q}{\partial Q^{j'} \partial P^j}} \widehat{\frac{\partial^{j+j'} p}{\partial Q^j \partial P^{j'}}}\right] = i \hbar f(p(\hat{Q},\hat{P})).
\end{equation}

As a specific example, consider the case $\hat{p}\equiv \hat{P}$.
Then, we find
\begin{equation}
	[\hat q,\hat p]
	= i \hbar \widehat{\frac{\partial q}{\partial Q}} = i \hbar f(\hat{p}).
\end{equation}
This relation can be used to find the explicit expression of $\hat q = q (\hat Q,\hat p)$. Indeed, suppose for a while to deal with c-number quantities. We can easily see that
\begin{equation}
\label{fQ}
	f(p) = \frac{\partial q}{\partial Q} \qquad \Rightarrow \qquad q = f(p) Q, 
\end{equation}
where we have required $q(Q=0, p)=0$. Clearly, in coming back to operators, we need to assign an ordering prescription.
For example, one could choose one of the following
\begin{equation}
	\hat{q}_1 = f(\hat{p}) \hat{Q}, \qquad\,\,\,\,
	\hat{q}_2 =  \hat{Q} f(\hat{p}), \qquad\,\,\,\,	\hat{Q}_3 =  \frac{1}{2} \left[f(\hat{p}) \hat{Q} + \hat{Q} f(\hat{p})\right].
\end{equation}
For reasons that will appear clear below, henceforth we shall work in the representation in which the operator $\hat{p}$ is multiplicative. In this case, we find
\begin{equation}
	\hat{Q} = i \hbar \frac{\diff}{\diff p}, 
\end{equation}
which leads to the following relations
\begin{equation}
	\hat{q}_1 =\,  i \hbar f(p) \frac{\diff}{\diff p},\qquad\,\,\,\,
	\hat{q}_2 =\,    i \hbar \left[f(p) \frac{\diff}{\diff p} + \frac{\diff f(p)}{\diff p}\right], \qquad\,\,\,\,
	\hat{q}_3 =\,  i \hbar \left[f(p) \frac{\diff}{\diff p} +\frac{1}{2} \frac{\diff f(p)}{\diff p}\right]. \label{eqn:positions}
\end{equation}
The representation $\hat{q}_1$ can be recognized as the one used in Ref.~\cite{KMM}.
However, the other two are equally applicable to the case of the commutator in Eq.~\eqref{eqn:GUP}.

\subsection{Maximally localized states with a generalized position operator}

Let us now identify $\hat q$ and $\hat p$ with the position and momentum operators of our physical system. 
Based on Eq.~\eqref{eqn:positions},  we are going to introduce the most general position operator in momentum space as
\begin{equation}
    \hat{q} = i \hbar \left[f(p) \frac{\diff}{\diff p} + A \frac{\diff f(p)}{\diff p}\right] = i \hbar [f(p)]^{1-A} \frac{\diff}{\diff p} [f(p)]^{A}.
    \label{Qrep}
\end{equation}
Notice that the three cases in Eq.~\eqref{eqn:positions} correspond to $A=0$ ($\hat{q}_1$), $A=1$ ($\hat{q}_2$) and $A=\frac{1}{2}$ ($\hat{q}_3$).

The purpose of introducing a deforming function $f(\hat p)$ as in Eq.~\eqref{eqn:GUP} is to  accommodate a minimal position uncertainty mimicking the granular structure of spacetime. As discussed in \cite{KMM}, in this case it makes no physical sense to work in the position eigenbasis, since
the related matrix elements obviously cannot be featured with the usual direct physical interpretation about positions.
Nevertheless, information on position is still accessible by introducing the so-called Maximally Localized States (MLS), which by definition minimize the position uncertainty consistently with Eq.~\eqref{eqn:GUP}. 

To find the explicit form of the MLS within our framework, let us consider the  equation they satisfy.
In the momentum representation, this takes the form of a differential equation~\cite{KMM}
\begin{equation}
\label{MLS}
	\frac{\diff}{\diff p} \psi^{(A)}_{\langle q \rangle} (p) 
	= \frac{1}{f(p)} \left\{\frac{\langle q \rangle}{i \hbar} - A \frac{\diff f(p)}{\diff p} - \frac{\langle f(p) \rangle (p - \langle p \rangle)}{2 (\Delta p)^2}\right\} \psi^{(A)}_{\langle q \rangle}(p), 
\end{equation}
where we have denoted by $\psi^{(A)}_{\langle q \rangle}$ the
maximally localized state around the generic position $\langle q \rangle$ and corresponding to the factor ordering described by the parameter $A$.
It is worth noticing that this relation
generalizes the MLS equation of Refs.\cite{KMM,Bosso:2020aqm} to the $A\neq 0$ case.

The solution of Eq.~\eqref{MLS} can be written, up to a normalization constant, as
\begin{equation}
	\psi_{\langle q \rangle}^{(A)}(p) = \chi^{(A)}(p) \exp\left[- i \frac{\langle q \rangle \, p_0(p)}{\hbar}\right], \label{eqn:mls}
\end{equation}
with 
\begin{equation}
\label{auxmom}
    p_0(p) =  \int \frac{\diff p}{f(p)}, \qquad\,\,\,\,
    \chi^{(A)}(p)	=  \exp \left[ - \frac{\langle f(p) \rangle}{2 (\Delta p)^2} \int \diff p \frac{p - \langle p \rangle}{f(p)}\right] [f(p)]^{-A}.
\end{equation}
Strictly speaking, the MLS
are obtained by further implementing the condition of absolutely smallest uncertainty  into Eq.~\eqref{eqn:mls}. Clearly, this condition is subjected to the specific form of $f(\hat p)$. For instance, for $f(\hat p)=1+\epsilon \hat p^2$ (which is the case considered in \cite{KMM,Bosso:2020aqm}), the MLS  are obtained by demanding $\langle \hat p\rangle=0$ and $\Delta p=1/\sqrt{\epsilon}$, with $\epsilon>0$  having dimensions of an inverse momentum squared.

Now, to make the comparison with Refs. \cite{KMM,Bosso:2020aqm} easier, it is convenient to recast $\chi^{(A)}(p)$ as
\begin{equation}
    \chi^{(A)}(p) = \chi^{(0)}(p) [f(p)]^{- A},
    \label{eqn:chi_A}
\end{equation}
where 
\begin{equation}
    \chi^{(0)}(p) = \exp \left[- \frac{\langle f(p) \rangle}{2 (\Delta p)^2} \int \diff p \frac{p - \langle p \rangle}{f(p)} \right]. 
\end{equation}
It is then straightforward to see that 
\begin{equation}
\psi^{(0)}_{\langle q\rangle}(p)= \chi^{(0)}(p)\exp\left[- i \frac{\langle q \rangle \, p_0(p)}{\hbar}\right],   
\end{equation}
are exactly the MLS found in Refs. \cite{KMM,Bosso:2020aqm} for $A=0$.

\subsection{Measure}

Using the same argument as in Ref.~\cite{KMM}, we are now interested in looking for the correct measure factor for the momentum space so as to make $\hat{q}$ symmetric.
In fact, we notice that neither $\hat{q}_1$ nor $\hat{q}_2$ in Eq.~\eqref{eqn:positions} are symmetric with the standard measure of momentum space, while $\hat{q}_3$ is such.
Thus, using a generic factor $g(p)\in\mathbb{R}$ for the new measure and letting the operators act on the dense domain $S_\infty$ of functions decaying faster
than any power \cite{KMM}, we have
\begin{eqnarray}
\nonumber
    \int_{-\infty}^{\infty}\diff p ~ g(p) \phi_1^\star (p) \hat{q} \phi_2(p)
    &\hspace{-0.2cm}=\hspace{-0.2cm}& i \hbar \int_{-\infty}^{\infty}\diff p ~ g(p) \phi_1^\star (p) \left[f(p) \frac{\diff}{\diff p} + A \frac{\diff f(p)}{\diff p}\right] \phi_2(p)\\[2mm]
&\hspace{-0.2cm}=\hspace{-0.2cm}& 
    \int_{-\infty}^{\infty} \diff p ~ \left\{ - i \hbar A g(p) \frac{\diff f(p)}{\diff p} \phi_1(p) + i \hbar \frac{\diff}{\diff p} \left[g(p) f(p) \phi_1(p)\right]\right\}^\star \phi_2(p), 
\end{eqnarray}
where we have performed a partial integration. 
Then, we demand that
\begin{equation}
    - i \hbar A g(p) \frac{\diff f(p)}{\diff p} \phi_1(p) + i \hbar \frac{\diff}{\diff p} \left[g(p) f(p) \phi_1(p)\right] = g(p) i \hbar \left[f(p) \frac{\diff}{\diff p} + A \frac{\diff f(p)}{\diff p}\right] \phi_1(p).
\end{equation}
With the further condition that for small $p$ we need to obtain back standard QM, \emph{i.e.} $g(0)=1$, this implies
\begin{equation}
    g(p) =  [f(p)]^{2 A - 1}.
\end{equation}
Notice that for $A=0$, we obtain the measure used in Refs.~\cite{KMM,Bosso:2020aqm}.
Furthermore,  for $A=\frac{1}{2}$ we have $g(p)=1~\forall p$.
This result is consistent with $\hat{q}_3$ being a symmetric operator with the standard measure of momentum space.

In terms of the momentum eigenstates $|p\rangle$, the identity operator can thus be expanded as
\begin{equation}
\label{ident}
\mathds{1}=\int_{-\infty}^{\infty}dp~[f(p)]^{2 A - 1}|p\rangle\langle p|,
\end{equation}
which indeed recovers the completeness relation used in Ref.~\cite{KMM} for $A=0$.

For practical purposes, it is worth noticing that the wavefunctions associated with different ordering prescriptions, \emph{i.e.} with different values of $A$, are related to each other.
In fact, consider the stationary Schr\"odinger equation for an arbitrary potential $U(\hat q)$ 
\begin{equation}
    \frac{p^2}{2m} \phi^{(A)}(p) + (U(\hat{q}) - E) \phi^{(A)}(p) = 0. \label{eqn:schrodinger}
\end{equation}
Since an arbitrary power of the position operator $\hat{q}$ can be written as
\begin{equation}
    (\hat{q})^n \phi^{(A)}(p) = \frac{(i \hbar)^n}{[f(p)]^{A}} \left( f(p) \frac{\diff}{\diff p}\right)^n [f(p)]^{A} \phi^{(A)}(p),
\end{equation}
when $U(\hat{q})$ can be written as a power series in $\hat{q}$, we can define the function $\phi^{(0)}(p) = [f(p)]^{A} \phi^{(A)}(p)$ with respect to which Eq.~\eqref{eqn:schrodinger} becomes
\begin{equation}
    \frac{p^2}{2m} \phi^{(0)}(p) + (U(\hat{q}_1) - E) \phi^{(0)}(p) = 0.
\end{equation}
Given the solution of such equation, one can then reconstruct the wavefunction associated to any specific ordering.
Thus, the wavefunctions associated to the various ordering are correlated.
Furthermore, this allows us to identify the corresponding Hilbert space.
In fact, how we saw above, since for $A = 1/2$ the measure of momentum space is unchanged, the corresponding Hilbert space is the usual space of $L^2(a,b)$ functions on a given (eventually infinite) domain $(a,b)$.
However, by the definition of the function $\phi^{(0)}(p)$ we also get that
\begin{equation}
    \phi^{(0)}  
    = {[f(p)]}^{1/2} \phi^{(1/2)}(p) \qquad \Rightarrow \qquad \phi^{(A)}(p) = [f(p)]^{1/2-A} \phi^{(1/2)}(p).
    \label{eqn:wf_conversion_order}
\end{equation}
Thus, given a particular ordering prescription $A$, the Hilbert space of the corresponding physical states $\phi^{(A)}(p)$ is $\mathcal{H} = \{\phi^{(A)}(p) : [f(p)]^{A-1/2} \phi^{(A)}(p) \in L^2\}$. 

\subsection{Momentum and (quasi-)position representations}
\label{sec:transforms}

We now proceed with the definition of an integral transform which is consistent with the deformed commutator \eqref{eqn:GUP} and has
the function \eqref{eqn:mls} as kernel.
By use of Eq.~\eqref{ident}, one can show that 
\begin{subequations}
	\begin{equation}
		\left[\mathcal{T}^{(A)}\right]^{-1}[\phi](\xi) 
		=  \frac{1}{\sqrt{2 \pi \hbar}} \int_{-\infty}^{\infty} \diff p ~ [f(p)]^{2A-1} ~ \psi^{(A)}{}^\star (p,\xi) \phi(p),\label{eqn:anti-transform}
		\end{equation}
\begin{equation}
	\hspace{14mm}	\mathcal{T}^{(A)}[\phi](p) 
		=  \frac{1}{\sqrt{2 \pi \hbar}} \int_{-\infty}^{\infty} \diff \xi ~ [f(p)]^{- 2A} \left[\psi^{(A)}{}^\star (p,\xi)\right]^{-1} \phi(\xi). \label{eqn:transform}
	\end{equation} \label{eqn:transforms}
\end{subequations}
Here $\psi^{(A)}(p,\xi) \equiv \psi^{(A)}_{\langle q \rangle}(p)$ and the parameter $\langle q \rangle \equiv \xi$ is now treated as a variable. Notice that the transformations \eqref{eqn:anti-transform} and \eqref{eqn:transform} are the inverse of each other, as it should be. Furthermore, 
$\mathcal{T}^{(A)}[\phi]$ is related to the corresponding transformation for $A=0$ defined in Ref.~\cite{Bosso:2020aqm} by
\begin{equation}
\label{TaT0}
    \mathcal{T}^{(A)}[\phi]=[f(p)]^{-A}\mathcal{T}^{(0)}[\phi].
\end{equation}

An useful relation for the subsequent discussion involves the product of two transformed function.
Specifically, we find
\begin{equation}
    \mathcal{T}^{(A)}[\phi_1]^\star (p) \mathcal{T}^{(A)}[\phi_2] (p) 
    = \frac{1}{\sqrt{2 \pi \hbar}} \frac{[f(p)]^{- 2A}}{\chi^{(A)}(p)} \mathcal{T}^{(A)} [\phi_1 \star \phi_2](p),
    \label{convo}
\end{equation}
where $(\phi_1 \star \phi_2)(\Xi)$ stands for the cross-correlation of the two functions, \emph{i.e.}
\begin{equation}
    (\phi_1 \star \phi_2)(\Xi) = \int_{-\infty}^\infty \diff \xi ~ \phi_1^\star(\xi) \phi_2(\Xi+\xi).
\end{equation}

With these integral transforms, we can now find the quasi-position representation of the various operators.
Specifically, it follows that the quasi-position representation of the momentum operator is $\hat{p}=p(\hat p_0)$, with $p(p_0)$ being the inverse function of $p_0(p)$ and
\begin{equation}
    \hat p_0=-i\hbar \frac{\diff}{\diff \xi}.
\end{equation}
On the other hand, by using the integral transform \eqref{eqn:anti-transform}, the representation of the position operator in the new space reads
\begin{equation}
\label{posope}
    \hat{q}=\xi + i\hbar\frac{\langle f(p)\rangle}{2\Delta p^2}\left(\hat p-\langle p\rangle\right), 
\end{equation}
that is clearly independent of the particular value of $A$.
Moreover, we can find that the momentum eigenfunction in quasi-position space is
\begin{equation}
	\phi_{\tilde{p}}(\xi) 
	= \left[\mathcal{T}^{(A)}\right]^{-1}[\delta(\tilde{p} - p)](\xi)
	= \frac{1}{\sqrt{2 \pi \hbar}} [f(\tilde{p})]^{2A-1} \chi^{(A)}(\tilde{p}) \exp\left[i \frac{\xi p_0(\tilde{p})}{\hbar}\right]. \label{eqn:momentum_eigen_position}
\end{equation}

Using the integral transforms defined above, the scalar product of states in terms of the quasi-position wave functions can be written as
\begin{equation}
    \langle\phi_1|\phi_2\rangle 
    = \int_{-\infty}^{\infty}dp ~ [f(p)]^{2A-1} \mathcal{T}^{(A)}[\phi_1]^\star(p)\mathcal{T}^{(A)}[\phi_2](p)
    = \frac{1}{\sqrt{2\pi\hbar}}\int_{-\infty}^{\infty}dp ~ \frac{1}{f(p) \chi^{(A)}(p)} \mathcal{T}^{(A)} [\phi_1 \star \phi_2](p),
    \label{scpr}
\end{equation}
where we have used the completeness relation in Eq.~\eqref{ident} along with Eq.~\eqref{convo}. 
Again, by employing Eqs.~\eqref{eqn:chi_A} and \eqref{TaT0}, it follows that the scalar product does not depend on the specific choice of $A$, as it was expected.

For later convenience, we specialize the above transforms to the case of the  quadratic model $f(p) = 1 + \epsilon p^2$ considered in Refs.~\cite{KMM}. 
Furthermore, we set $A=0$, the other orderings being recovered via the prescription \eqref{TaT0}. 
In this context, the transforms in Eqs.~\eqref{eqn:transforms} become
\begin{eqnarray}
    \mathcal{T}^{-1}[\phi](\xi) &\hspace{-0.2cm}=\hspace{-0.2cm} & \frac{1}{\sqrt{2\pi\hbar}} \int_{-\infty}^{\infty} \frac{dp}{(1+\epsilon p^2)^{3/2}} \exp \left[i \frac{\xi \arctan\left(\sqrt{\epsilon}p\right)}{\sqrt{\epsilon} \hbar}\right] \phi(p),\\[2mm]
    \mathcal{T}[\phi](p) &\hspace{-0.2cm}=\hspace{-0.2cm} & \frac{1}{\sqrt{2\pi\hbar}} \int_{-\infty}^{\infty} d \xi~ \sqrt{1+\epsilon p^2} \exp \left[- i \frac{\xi \arctan\left(\sqrt{\epsilon}p\right)}{\sqrt{\epsilon} \hbar}\right] \phi(\xi).
\end{eqnarray}
An interesting feature of this GUP  can be highlighted by rewriting the anti-transform $\mathcal{T}^{-1}$ in terms of the the auxiliary momentum $p_0$ introduced in Eq.~\eqref{auxmom}.
First, we notice that the relation between $p$ and $p_0$ in the present case is
\begin{equation}
\label{pp0}
    p=\frac{1}{\sqrt{\epsilon}} \tan(\sqrt{\epsilon}p_0).
\end{equation}
Therefore, we have
\begin{eqnarray}
\nonumber
    \mathcal{T}^{-1}[\phi](\xi)
   & \hspace{-0.2cm}= \hspace{-0.2cm}& \frac{1}{\sqrt{2\pi\hbar}} \int_{-\frac{\pi}{2\sqrt{\epsilon}}}^{\frac{\pi}{2\sqrt{\epsilon}}} dp_0 ~ \cos(\sqrt{\epsilon}p_0) e^{i \frac{\xi p_0}{\hbar}} \phi(p(p_0))\\[2mm]
   &\hspace{-0.2cm} =\hspace{-0.2cm}& \frac{1}{\sqrt{8\pi\hbar}} \int_{-\frac{\pi}{2\sqrt{\epsilon}}}^{\frac{\pi}{2\sqrt{\epsilon}}} dp_0 ~ \left[e^{i\left(\frac{\xi}{\hbar} + \sqrt{\epsilon}\right) p_0} + e^{i\left(\frac{\xi}{\hbar} - \sqrt{\epsilon}\right) p_0}\right] \phi(p(p_0)).
    \label{eqn:atrans_p0}
\end{eqnarray}
It is worth observing that the anti-transform as described by the last expression resembles the sum of two transforms producing functions with argument $\frac{\xi}{\hbar} \pm \sqrt{\epsilon}$.
A better insight can be obtained considering momentum eigenfunctions, that is $\phi(p_0) = \delta(p_0-\tilde{p}_0)$, with $\tilde{p}_0$ associated to the momentum eigenvalue related with such function.
In this case, the anti-transform produces two plane waves, both with wavenumber $k = p_0/\hbar$, with a phase difference $\Delta \varphi = 2\sqrt{\epsilon} p_0$. 
This result is a direct consequence of the minimal length framework considered in this paper.
We can interpret it by saying that the combination of the two functions is in fact the object under study (in the example, the momentum eigenstate) and that the two components cannot be resolved because of the existence of a minimal uncertainty in position. Clearly, the standard single-wave picture is recovered in the limit $f(p)\rightarrow\ide$, \emph{i.e.} $\epsilon\rightarrow0$.

\section{GUP-deformed harmonic oscillator}
\label{Analytic}
In this Section we solve the linear harmonic oscillator (HO) in the presence of a generalized commutator like that in Eq.~ \eqref{eqn:GUP}. We 
revisit both the analytic and algebraic methods, deriving corrections to the eigenfunctions and energy spectrum. In particular, the algebraic approach allows us to understand how the Fock space structure is affected when deforming the fundamental algebra.  This finds non-trivial applications in the study of minimal length effects on  oscillator frequency measurements, as we shall discuss below. 

\subsection{Analytic method}
Let us start from the stationary Schr\"odinger equation for the linear HO written for $A$=0 (any other ordering can be recovered from this as discussed above). This reads
\begin{equation}
    \frac{\diff}{\diff p} \left[f(p) \frac{\diff}{\diff p}\psi(p)\right] + \frac{2}{\hbar^2 m \omega^2} \frac{1}{f(p)} \left(E - \frac{p^2}{2m}\right) \psi(p) = 0.
    \label{diffequ}
\end{equation}
Inspired by previous works in the literature, here we consider a polynomial deformation of the commutator in the form
\begin{equation}
\label{deforfunc}
    f(p) = 1 - 2 \delta p + (\delta^2 + \epsilon) p^2.
\end{equation}
In order for the r.h.s. of Eq.~\eqref{eqn:GUP} to be non-vanishing, we require $\epsilon>0$.
Furthermore, for dimensional reasons we assume that $\delta$
goes as the inverse of the characteristic energy-scale where corrections to the HUP  become relevant\footnote{In the effective quantum description of gravity, this scale is provided by the Planck mass $m_p$, so that $\delta=\delta_0/(m_p c)$ and $\epsilon=\epsilon_0/(m_p c)^2$, with both $\delta_0$ and $\epsilon_0$ dimensionless parameters of order unity.}.

It is a matter of trivial calculations to show that the GUP~\eqref{eqn:GUP} with the above deforming function  still admits a minimal position uncertainty, the MLS being now identified by the conditions $\langle \hat p\rangle=\delta/(\delta^2+\epsilon)$ and $\Delta p= \sqrt{\epsilon}/(\delta^2+\epsilon)$.  
Clearly, the framework of Ref.~\cite{KMM} with a quadratic GUP model and the ordinary quantum mechanics
are recovered for $\delta\rightarrow0$ and $\delta,\epsilon\rightarrow0$,
respectively.

Since the coefficients of the differential equation~\eqref{diffequ} are rational functions of $p$ and since it has three regular singular points (two of which lying at finite values of $p$ and the other at $p\rightarrow\infty$), it is possible to recast the same equation in the form~\cite{HGF_thesis}
\begin{equation}
    \zeta (1 - \zeta) \frac{\diff^2}{\diff \zeta^2} f(\zeta) + [c - (a+b+1)\zeta] \frac{\diff}{\diff \zeta} f(\zeta) - a b f(\zeta) = 0, \label{eqn:hf_equation}
\end{equation}
whose solution can be written in general as
\begin{equation}
\label{Ghf}
    f(\zeta) = \zeta^{\rho'} (1-\zeta)^{\sigma'} {}_2F_1(a',b';c';\zeta'),
\end{equation}
with $\rho'$, $\sigma'$, $a'$, $b'$, $c'$ linear functions of the parameters $a$, $b$, $c$ and $\zeta'$ a linear fractional function of $\zeta$ to be determined. 
Here, ${}_2F_1(a',b';c';\zeta')$ is the Gauss hypergeometric series.

In order to find the solution of Eq.~\eqref{diffequ} explicitly, it is useful to shift the two finite singularities toward the two reference values $0$ and $1$. This can be done by observing that $f(p)$ has two roots
\begin{equation}
    p_\pm = \frac{\delta \pm i \sqrt{\epsilon}}{\delta^2 + \epsilon}.
\end{equation}
We can then define a new variable 
\begin{equation}
    \zeta = \frac{p - p_-}{p_+ - p_-} = \frac{1}{2} \left[ 1 - i \frac{(\delta^2 + \epsilon) p - \delta}{\sqrt{\epsilon }}\right],
\end{equation}  
such that 
\begin{equation} 
p = \frac{\delta + i \sqrt{\epsilon}(2 \zeta - 1)}{\delta^2 + \epsilon}.
\end{equation}
In this way, the Schr\"odinger equation becomes
\begin{equation}
    \zeta (1 - \zeta) \frac{\diff^2 \psi}{\diff \zeta^2} + (1 - 2 \zeta) \frac{\diff \psi}{\diff \zeta} + \frac{\left[\delta - i (1 - 2 \zeta) \sqrt{\epsilon }\right]^2 - 2 E m \left(\delta ^2+\epsilon \right)^2}{4 (1 - \zeta) \zeta m^2 \omega ^2 \hbar ^2 \epsilon \left(\delta ^2+\epsilon \right)^2} \psi(\zeta) = 0, 
\end{equation}
whose finite singularities are indeed located at $\xi=0$ and $\xi=1$. This equation cannot yet be compared with Eq.~\eqref{eqn:hf_equation}, since the term multiplying  $\psi(\zeta)$ depends explicitly on $\zeta$, while in~\eqref{eqn:hf_equation} it does not.

Let us then consider a test function in the form
\begin{equation}
    \psi(\zeta) = \zeta^\rho (1-\zeta)^\sigma \phi(\zeta).
\end{equation}
The Schr\"odinger equation reads
\begin{eqnarray}
\nonumber
 &&   \zeta (1 - \zeta) \frac{\diff^2 \phi}{\diff \zeta^2} + [1 + 2 \rho - 2 \zeta  (\rho + \sigma +1)] \frac{\diff \phi}{\diff \zeta} \\[2mm]
 &&  \hspace{2mm} + \frac{4 \left[(\rho + \sigma) (\rho + \sigma + 1) \zeta^2 - (2 \rho + 1) (\rho + \sigma) \zeta + \rho^2 \right] + \frac{\left[\delta - i \sqrt{\epsilon} (1-2 \zeta ) \right]^2 - 2 E m (\delta^2 + \epsilon)^2}{m^2 \omega ^2 \epsilon  \hbar ^2 \left(\delta ^2+\epsilon \right)^2}}{4 (1 - \zeta) \zeta } \phi (\zeta) = 0. \label{eqn:schrodinger_zeta}
\end{eqnarray}
As it stands,  the last term  still depends on $\zeta$. However, we can now set the parameters $\rho$ and $\sigma$ so as to trivialize such dependence. 
Specifically, we can choose
\begin{equation}
    \rho = \pm \frac{\sqrt{2 E m - \frac{1}{\left(\delta +i \sqrt{\epsilon }\right)^2}}}{2 m \omega  \hbar \sqrt{\epsilon } }, \qquad\,\,\,\,
    \sigma = \pm \frac{\sqrt{2 E m - \frac{1}{\left(\delta - i \sqrt{\epsilon }\right)^2}}}{2 m \omega  \hbar \sqrt{\epsilon } }. \label{eqn:rho_sigma}
\end{equation}
It is worth noticing that both $\rho$ and $\sigma$ are in general complex numbers, the one being the conjugate of the other if chosen with the same sign. However, they become real for vanishing  $\delta$, in which case they reduce to
the parameters $\alpha_1$, $\alpha_2$, $\beta_1$, and $\beta_2$ of Ref.~\cite{KMM}, as it should be.  
Furthermore, in the limit $\epsilon\rightarrow0$ they are real only when $\delta^{-2} < 2 E m$, which means that $\delta^{-1}$ should be (in module) smaller than the characteristic momentum $\sqrt{2 E m}$ of a quantum harmonic oscillator.
Were this true, we should have already observed the implications of a minimal length (see Eq.~\eqref{deforfunc}).
Thus, we see that either $\delta = 0$ or $\rho$ and $\sigma$ must necessarily be complex quantities.
In what follows, we start by considering the $\delta\neq0$ case. 

Going back to Eq.~\eqref{eqn:schrodinger_zeta} and comparing it to Eq.~\eqref{eqn:hf_equation}, we can also identify
\begin{equation}
    c = 1 + 2 \rho, \qquad\,\,\,\, a + b + 1 = 2 (\rho + \sigma + 1),
        \label{abrelat}
\end{equation}
and
\begin{equation}
    - a b = \frac{4 \left[(\rho + \sigma) (\rho + \sigma + 1) \zeta^2 - (2 \rho + 1) (\rho + \sigma) \zeta + \rho^2 \right] + \frac{\left[\delta - i \sqrt{\epsilon} (1-2 \zeta ) \right]^2 - 2 E m (\delta^2 + \epsilon)^2}{m^2 \omega ^2 \epsilon  \hbar ^2 \left(\delta ^2+\epsilon \right)^2}}{4 (1 - \zeta) \zeta},
\end{equation}
recalling that this last expression is independent of $\zeta$ for the above choice  of $\rho$ and $\sigma$.
We then find
\begin{eqnarray}
\label{acoef}
    a 
    &\hspace{-0.2cm}=\hspace{-0.2cm} & \frac{1}{2} \left[1 + 2 \rho + 2 \sigma \mp \frac{\sqrt{m^2 \omega ^2 \hbar ^2 \left(\delta^2 + \epsilon \right)^2 + 4}}{m \omega  \hbar \left(\delta ^2+\epsilon \right)}\right],\\[1mm]
    \label{bcoef}
    b 
   &\hspace{-0.2cm} = \hspace{-0.2cm}& \frac{1}{2} \left[1 + 2 \rho + 2 \sigma \pm \frac{\sqrt{m^2 \omega ^2 \hbar ^2 \left(\delta^2 + \epsilon \right)^2 + 4}}{m \omega  \hbar \left(\delta ^2+\epsilon \right)}\right],\\[1mm]
    c &\hspace{-0.2cm}=\hspace{-0.2cm} & 1 + 2 \rho.
\end{eqnarray}
With the above settings, the solution $\psi(\zeta)$ is given in terms of Gauss hypergeometric series as~\cite{HGF_Bailey}
\begin{equation}
\label{psizita}
    \psi(\zeta) = \mathcal{A} \psi_1(\zeta) + \mathcal{B} \psi_2(\zeta),
\end{equation}
where
\begin{eqnarray}
\nonumber
     \psi_1(\zeta) &\hspace{-0.2cm} =\hspace{-0.2cm} & \zeta^\rho (1-\zeta)^\sigma {}_2F_1(a,b;c;\zeta) \nonumber\\[1mm]
\label{psi1}
    &\hspace{-0.2cm} =\hspace{-0.2cm}  & \left\{\frac{\left(\delta^2 + \epsilon\right) \left[1 - 2 \delta p + (\delta^2 + \epsilon) p^2\right]}{4 \epsilon}\right\}^{\rho} \left\{\frac{1}{2} - i \frac{\delta - \left(\delta^2 + \epsilon\right) p}{2 \sqrt{\epsilon }}\right\}^{\sigma - \rho} {}_2F_1 \left(a,b;c; \frac{1}{2} + i \frac{\delta - \left(\delta^2 + \epsilon\right) p}{2 \sqrt{\epsilon }}\right),\\[2.5mm]
\nonumber
    \psi_2(\zeta) &\hspace{-0.2cm} =\hspace{-0.2cm}  & \zeta^\rho (1-\zeta)^\sigma \zeta^{1-c} {}_2 F_1 (a+1-c,b+1-c;2-c;\zeta)\\[1mm]
\nonumber
    &\hspace{-0.2cm} =\hspace{-0.2cm}  & \left\{\frac{\left(\delta^2 + \epsilon\right) \left[1 - 2 \delta p + (\delta^2 + \epsilon) p^2\right]}{4 \epsilon}\right\}^{-\rho} \left\{\frac{1}{2} - i \frac{\delta - \left(\delta^2 + \epsilon\right) p}{2 \sqrt{\epsilon }}\right\}^{\sigma + \rho} \\[1mm]
    && \times {}_2F_1 \left(a+1-c,b+1-c;2-c; \frac{1}{2} + i \frac{\delta - \left(\delta^2 + \epsilon\right) p}{2 \sqrt{\epsilon }}\right),
    \label{psi2}
\end{eqnarray}
and $\mathcal{A}$ and $\mathcal{B}$ are two constants.

A remark is now in order. We know that the quantum mechanical HO eigenfunctions are given by the product
of a Gaussian factor by Hermite
polynomials. In order to recover this behavior for vanishing deformations parameters, we look for GUP solutions in the cases where the hypergeometric series reduces to a polynomial. This occurs whenever one of the first two entries of the hypergeometric series is a negative integer. Then, for $\psi_1$ to be acceptable, it must be verified that either $a$ or $b$ is equal to $-n$, where $n\in \mathbb{N}$ will be later identified with the energy level of HO states.
In turn, this requires that $\rho$ and $\sigma$ in Eq.~\eqref{eqn:rho_sigma} must be chosen with the same sign, as this is the only way for $a$ and $b$
to have vanishing imaginary parts (indeed, we recall that $\rho$ and $\sigma$ are complex conjugates if they are chosen with the same sign).
On the other hand, from Eq.~\eqref{psi2} the solution $\psi_2$ gives back the correct limit if either $a+1-c$
or $b+1-c$ is a negative integer.
For the same reasons as above and since $c\in\mathbb{C}$, $\rho$ and $\sigma$ must now have opposite signs.
It is thus clear that
the consistency with QM forces 
us to select either $\mathcal{A}=0$ or $\mathcal{B}=0$ in Eq.~\eqref{psizita}. Without loss of generality, we focus on the latter case, though it is possible to show that the other choice leads to the same physical predictions.

Let us then assume $\psi(\zeta)=\mathcal{A}\psi_1(\zeta)$, with the condition that $\rho$ and $\sigma$
have the same sign.
In order to study the asymptotic behavior of $\psi$, it is worth analyzing the two terms multiplying the hypergeometric function in Eq.~\eqref{psi1} separately.
As for the first term, since $\rho \in \mathbb{C}$, it is characterized by an oscillating part, due to $\Im(\rho)$, and a part evolving as a power of $1 - 2 \delta p + (\delta^2 + \epsilon) p^2$ with exponent $\Re(\rho)$. As for the second term, we can write it as
\begin{equation}
    \left\{\frac{1}{2} - i \frac{\delta - \left(\delta^2 + \epsilon\right) p}{2 \sqrt{\epsilon }}\right\}^{\sigma - \rho}
    = \exp \left\{ i \Im(\sigma) \ln \frac{[\delta - (\delta^2 + \epsilon) p]^2 + \epsilon}{4 \epsilon}\right\} \exp \left\{ 2 \Im(\sigma) \arctan \left[\frac{\delta - (\delta^2 + \epsilon) p }{\sqrt{\epsilon }}\right]\right\},
\end{equation}
where we used the relations
\begin{equation}
    \sigma - \rho =  2 i \Im(\sigma), \qquad\,\,\,\,
    \frac{1}{2} - i \frac{\delta - \left(\delta^2 + \epsilon\right) p}{2 \sqrt{\epsilon }} = \frac{1}{2} \sqrt{\frac{[\delta - (\delta^2 + \epsilon) p]^2 + \epsilon}{\epsilon }} \exp \left\{i \arctan \left[\frac{- \delta + (\delta^2 + \epsilon) p }{\sqrt{\epsilon }}\right]\right\}.
\end{equation}
Therefore, this term consists of an oscillating part modulated in amplitude by a monotonic term which for $p\rightarrow\pm\infty$ tends to the finite quantity $\exp {\left[\mp \pi \Im(\sigma)\right]}$.
Altogether, 
the requirement that $\psi(\zeta)$
does not diverge at infinity implies that 
$\Re(\rho)$ (and, thus, $\Re(\sigma))$ must be negative, which means
that Eq.~\eqref{eqn:rho_sigma} should be  considered with negative signs.\footnote{Notice that, if we choose $\psi(\zeta)=\mathcal{B}\psi_2(\zeta)$, 
the convergence for large $p$ leads to the conditions $\Re(\rho)>0$ and $\Re(\sigma)<0$.}    
From Eq.~\eqref{abrelat}, we then demand that 
\begin{equation}
    4 \Re(\rho) = \Re(a + b - 1) = a + b - 1 < 0\,,
    \label{Rerho}
\end{equation}
where we have exploited the fact that  $a$ and $b$ are both real in the present setting. 

As discussed above, let us now impose 
the condition that either $a$ or $b$ is a negative integer. For instance, by choosing
$a = -n$, we get
\begin{equation}
\label{abcond}
    b = a \pm \frac{\sqrt{m^2 \omega ^2 \hbar ^2 \left(\delta^2 + \epsilon \right)^2 + 4}}{m \omega  \hbar \left(\delta ^2+\epsilon \right)} = - n \pm \frac{\sqrt{m^2 \omega ^2 \hbar ^2 \left(\delta^2 + \epsilon \right)^2 + 4}}{m \omega  \hbar \left(\delta ^2+\epsilon \right)}.
\end{equation}
Thus, we will consider the bottom sign so that the condition~\eqref{Rerho} is satisfied for every value of $n$.
Alternatively, one can impose $b = - n$ and choose the top sign for $a$.
In fact, these two choices are equivalent since the solution is symmetric under exchange of $a$ and $b$. 

Now, the conditions~\eqref{abcond} allow us to find the energy spectrum.
In fact, since $a + b = 1 + 2 (\rho + \sigma)$, using Eq.~\eqref{eqn:rho_sigma} with the minus sign for both $\rho$ and $\sigma$, we find that for the $n$-th level
\begin{eqnarray}
\nonumber
E_n^{(GUP)}&\hspace{-0.5cm}=\hspace{-0,5cm}& \frac{\omega  \hbar \left(\frac{1}{2} + n\right) \sqrt{4 + \left(\delta^2 + \epsilon\right)^2 m^2 \omega^2 \hbar^2}}{2(\delta^2 + \epsilon)} \left\{\frac{\delta ^2}{\left[1 - \left(\delta^2 + \epsilon\right)^2 m^2 \omega ^2 \hbar^2 n (1 + n)\right]^2}+\epsilon \right\}
    + \frac{\epsilon m \omega^2 \hbar^2}{4} \left(1 + 2 n + 2 n^2\right)\\[1mm]
  &&+ \frac{ \delta ^2 m \omega^2 \hbar^2}{4 \left[1 - \left(\delta^2 + \epsilon\right)^2 m^2 \omega ^2 \hbar^2 n (1 + n)\right]^2} \left[2 \left(\delta^2 + \epsilon\right)^2 m^2 \omega ^2 \hbar^2 n^2 (1 + n)^2 - \left(1 + 6 n + 6 n^2\right)\right], 
    \label{spectrumPas}
\end{eqnarray}
to be compared with the standard quantum mechanical result
\begin{equation}
    E^{(QM)}_n=\hbar\omega\left(n+\frac{1}{2}\right).
    \label{QMspec}
\end{equation}
One can easily check that the harmonic oscillator 
spectrum of Ref.~\cite{KMM} is recovered for $\delta\rightarrow0$, 
while the limit $\delta,\epsilon\rightarrow0$
gives back the standard result~\eqref{QMspec}.
Let us also remark that Eq.~\eqref{spectrumPas}
agrees, up to second order in the inverse Planck momentum, with the results obtained in~\cite{BossoCohe}.

\begin{figure}[t]
\centering
\includegraphics[angle=0,width=0.7\textwidth]{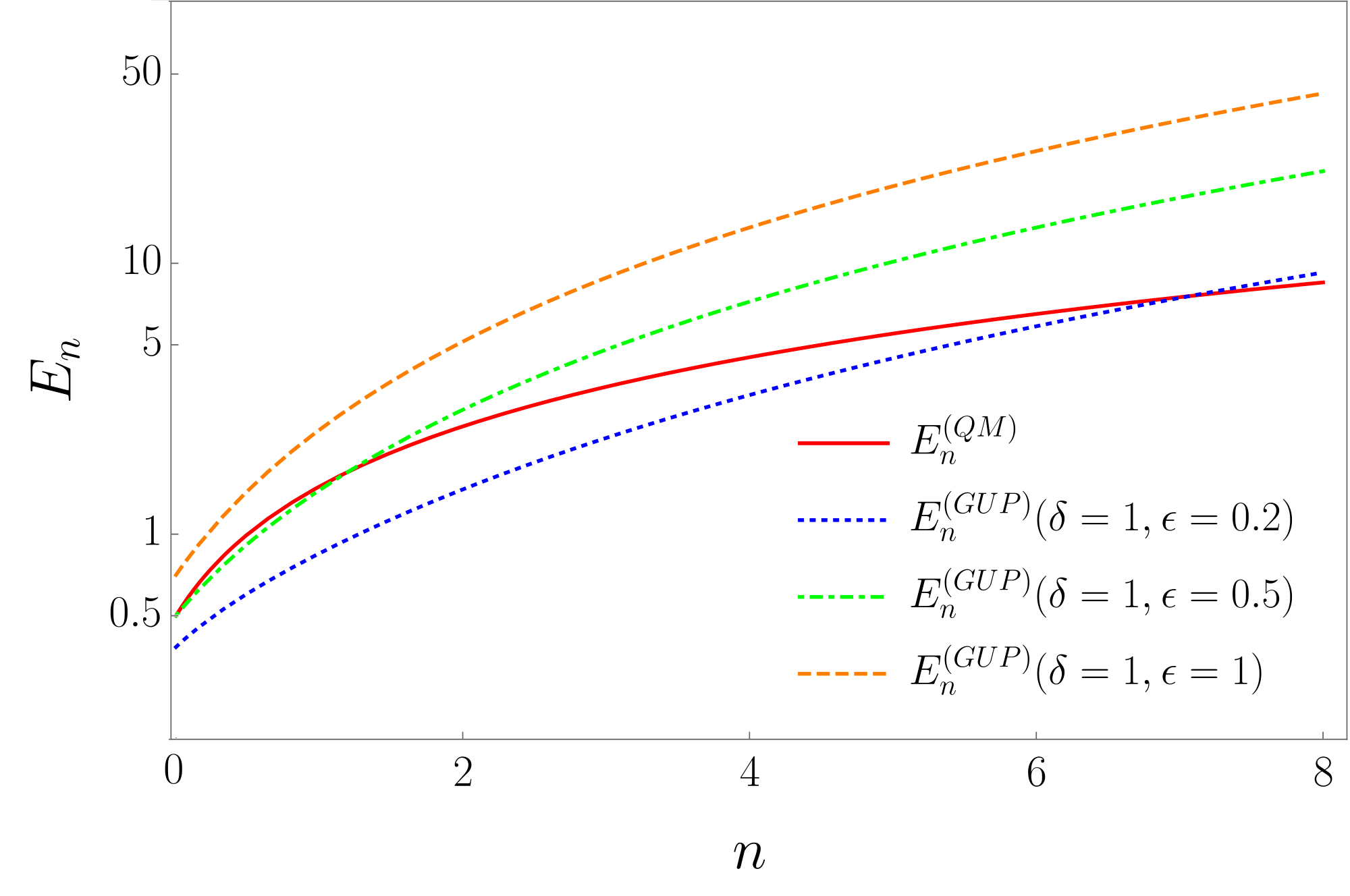}
\caption{Energy spectrum as a function of the level number $n$. We have set for simplicity  $\hbar=m=\omega=1$. The log-scale has been used on the vertical axis.}
\label{enerplot}
\end{figure}

Now, for large values of $n$, it is possible to show that the dominant term in the  spectrum~\eqref{spectrumPas} goes as
\begin{equation}
    E_n \simeq 
     \frac{\epsilon m \omega^2 \hbar^2}{2} n^2\,.
   \label{largensp}
\end{equation}
We see that, as long as $\epsilon > 0$, the energy increases quadratically with $n$, which is the same behavior found in~\cite{KMM}.
Thus, for $n$ large enough, the linear GUP-correction in Eq.~\eqref{deforfunc} does not affect the energy levels of the harmonic oscillator. 
Another remarkable feature
can be derived from the study of the zero-point energy. Indeed, by imposing that the modified vacuum energy $E^{(GUP)}_{n=0}$ 
is equal to the standard value $E_{n=0}^{(QM)}=\hbar\omega/2$, we obtain the following condition between the GUP parameters
\begin{equation}
\epsilon=\frac{\delta^2}{1+m\omega\hbar\delta^2}\,.
\end{equation}
Let us denote the r.h.s of this equation by $\bar{\epsilon}$. The above relation allows us to distinguish among three different regimes (see Fig.~\ref{enerplot}):
\begin{itemize}
    \item[-] for $\epsilon<\bar\epsilon$, we have $E_{n=0}^{(GUP)}<E_{n=0}^{(QM)}$. In this case, there exists only one intersection between the function~\eqref{spectrumPas} and the ordinary spectrum~\eqref{QMspec}. This is due to the fact that $E_n^{(GUP)}$ is a monotonically increasing function of the level number $n$ and, for $n$ large enough, it grows faster than the linear QM spectrum, according to Eq.~\eqref{largensp}. 
    \item[-] For $\epsilon=\bar{\epsilon}$, we have two intersection points, one of which at $n=0$. This occurs because, for small $n$, $E_{n}^{(QM)}$ grows faster than $E_{n}^{(GUP)}$, while this trend is reversed for increasing $n$, 
    as explained above.
    \item[-] A similar behavior occurs for $\epsilon\gtrsim\bar{\epsilon}$. However, in this case $E_{n=0}^{(GUP)}>E_{n=0}^{(QM)}$, which entails that both intersections lie at positive values of $n$. 
    \item[-] By further increasing $\epsilon$, the two curves have no intersection at all. Indeed, in this regime the growth rate of the QM spectrum is never high enough to fill the initial gap between $E_{n=0}^{(GUP)}$ and $E_{n=0}^{(QM)}$.
As a result, the modified spectrum lies above the QM prediction for any value of $n$.
\end{itemize}

Finally, let us focus on the characterization of the 
energy spacing between adjacent levels.
In QM it is well-known that the allowed energies are equally spaced and separated by a distance $\hbar\omega$.
Here the situation becomes far more complicated. The explicit expression of $\Delta E_n\equiv E^{(GUP)}_{n+1}-E^{(GUP)}_n$ is rather awkward to exhibit. However, an interesting property
can still be derived in the regime of large level numbers $n$, where the spacing is found growing linearly with $n$. This is consistent with the behavior of the spectrum exhibited in Eq.~\eqref{largensp} and implies that the higher $n$, the larger  the amount of energy that must be supplied to the harmonic oscillator to
push it to the next energy level. Qualitatively speaking, this result could be somehow expected, since
the r.h.s. of the deformed commutator~\eqref{eqn:GUP} is no longer a constant and increases for $p$ large enough. More quantitative details on the ladder structure of the spectrum will be given below by studying the algebraic solution of the harmonic oscillator.

\subsection{Algebraic method}
\label{Algebraic}
Above we imposed $a$ (or equivalently $b$) to be a negative integer number, which in turn corresponds to the energy quantum number.
Thus, it is clear that increasing $n$ by 1, which corresponds to climbing up the energy ladder, implies reducing $a$ and $b$ by 1.
On the other hand, since under our assumptions $c$ can be written as
\begin{equation}
c = -n+\frac{1}{2}\left(1-\frac{\sqrt{m^2 \omega ^2 \hbar ^2 \left(\delta^2 + \epsilon \right)^2 + 4}}{m \omega  \hbar \left(\delta ^2+\epsilon \right)}\right)+2 i \Im(\rho),
\end{equation}
increasing $n$ by 1 does not correspond to changing $c$ by an integer quantity, because of the $n$-dependence of $\Im(\rho)$. 
Given the properties of Gauss hypergeometric functions, this poses an issue: it is not possible to find an operator composed by multiplicative and derivative terms which, applied to any energy eigenfunction of quantum number $n$, leads us to another eigenfunction of quantum number $n\pm1$.
However, in agreement with~\cite{KMM}, imposing $\delta=0$ we have $\Im(\rho)=0$, allowing for a possible ladder operator.
It is worth noticing that is the only possibility to implement efficiently the algebraic method. In other terms, the presence of a first degree term in $p$ in the deforming function impedes the possibility of climbing the HO energy ladder.

Therefore, from now on we will restrict our description to the case $\delta = 0$. 
In particular, we have
\begin{eqnarray}
    \rho_n = \sigma_n &\hspace{-0.2cm}  =\hspace{-0.2cm}  & - \frac{n}{2} - \frac{1}{4} \left(1 + \sqrt{1 + \frac{4}{m^2 \omega^2 \epsilon^2 \hbar^2}}\right)
    = \rho_0 - \frac{n}{2},\\[1mm]
%
    a_n &\hspace{-0.2cm} =\hspace{-0.2cm}  & - n, \\[1mm]
    b_n &\hspace{-0.2cm} =\hspace{-0.2cm}  & - n - \sqrt{1 + \frac{4}{m^2 \omega^2 \epsilon^2 \hbar^2}}
    = 4 \rho_n + 1 - a_n, \\[1mm]
    c_n &\hspace{-0.2cm} =\hspace{-0.2cm}  & - n + \frac{1}{2}\left(1 - \sqrt{1 + \frac{4}{m^2 \omega^2 \epsilon^2 \hbar^2}}\right)
    = 1 + 2 \rho_n, \\[1mm]
    E_n &\hspace{-0.2cm} =\hspace{-0.2cm}  & \frac{\epsilon m \omega^2 \hbar^2}{2} \left[\left(\frac{1}{2} + n\right) \sqrt{1 + \frac{4}{m^2 \omega^2 \epsilon^2 \hbar^2}} + \frac{1}{2} \left(1 + 2 n + 2 n^2\right)\right]
    = \frac{\epsilon m \omega^2 \hbar^2}{2} \left[-4 \rho_0 \left(\frac{1}{2} + n\right) + n^2 \right],
    \label{Enspec}
\end{eqnarray}

By using the properties of Gauss hypergeometric functions, in particular
\begin{eqnarray}
    \left\{\left[1 - \frac{(a+b-1) \zeta}{c-1}\right] - \frac{ \zeta (\zeta-1)}{c-1} \frac{\diff}{\diff \zeta}\right\} \,_2F_1(a,b;c;\zeta ) &\hspace{-0.2cm} =\hspace{-0.2cm}  & \,_2F_1(a-1,b-1;c-1;\zeta), \\[2mm]
    \frac{c}{a b} \frac{\diff}{\diff \zeta} \,_2F_1(a,b;c;\zeta) &\hspace{-0.2cm} =\hspace{-0.2cm}  & \,_2F_1(a+1,b+1;c+1;\zeta ),
\end{eqnarray}
we can define the operators $A^\pm$ as\footnote{Since there is no ambiguity, henceforth we drop the hats from operators to simplify the notation.}
\begin{eqnarray}
\nonumber
    A^\pm &\hspace{-0.2cm} =\hspace{-0.2cm}  & \frac{\omega \hbar \sqrt{\epsilon m}}{2} \left[\pm \frac{1 - 2 \zeta}{\sqrt{\zeta(1-\zeta )}} \left(\rho_0 - \frac{N}{2}\right) + \sqrt{\zeta(1-\zeta )} \frac{\diff}{\diff \zeta } \right]\\[2mm]
    &\hspace{-0.2cm} =\hspace{-0.2cm}  & \frac{\omega \hbar \sqrt{\epsilon m}}{2} \left[\pm \frac{2 i p \sqrt{\epsilon }}{\sqrt{1 + \epsilon p^2}} \left(\rho_0 - \frac{N}{2}\right) + \frac{i \sqrt{1 + \epsilon p^2}}{\sqrt{\epsilon }} \frac{\diff}{\diff p}\right],
    \label{eqn:Apm_p}
\end{eqnarray}
where 
\begin{equation}
      N \psi_n (\zeta) = n \psi_n(\zeta),
\end{equation}
such that
\begin{equation}
    A^+ \psi_n (\zeta) =  \frac{\omega \hbar \sqrt{\epsilon m}}{2} (2 \rho_0 - n)\psi_{n+1} (\zeta), \qquad\,\,\,\,
    A^- \psi_n (\zeta) =  \frac{\omega \hbar \sqrt{\epsilon m}}{2} \frac{n (1 + 4 \rho_0-n)}{n - 2 \rho_0 - 1} \psi_{n-1} (\zeta). \label{eqn:ladderA}
\end{equation}
It is worth noticing that, up to a constant (dimensional) factor, such operators have the correct limit for $\epsilon \to 0$, that is,
\begin{equation}
    \lim_{\epsilon \to 0} A^+ =  \sqrt{\frac{\hbar \omega}{2}} a^\dagger, \qquad\,\,\,\,\lim_{\epsilon \to 0} A^- =  \sqrt{\frac{\hbar \omega}{2}} a,
\end{equation}
where $a$ and $a^\dagger$ are the usual ladder operators of standard quantum mechanics.
Furthermore, it is straightforward to see that
\begin{equation}
    A^+ A^- =  \frac{\omega^2 \hbar^2 \epsilon m}{4} N (N - 1 - 4 \rho_0), \qquad\,\,\,\,
    A^- A^+ = \frac{\omega^2 \hbar^2 \epsilon m}{4} (N+1) (N - 4 \rho_0). \label{eqn:products}
\end{equation}
Thus, we obtain
\begin{equation}
    [A^-, A^+] =  \frac{\omega^2 \hbar^2 \epsilon m}{2}  (N-2\rho_0), \qquad\,\,\,\,
    \{A^-,A^+\} =  \frac{\omega^2 \hbar^2 \epsilon m}{2} [N^2 - (2N+1) 2 \rho_0]
    = H,
    \label{commutatoreAmAp}
\end{equation}
where in the last equality we have compared the expression for the anti-commutator $\{A^-,A^+\}$ with Eq.~\eqref{Enspec}.
Such expressions resemble the canonical structure exhibited in quantum mechanics.
However, unlike the standard case, the commutator $[A^-,A^+]$, when acting on an energy eigenstate, now produces terms that depend on the quantum number $n$.
Therefore, such term can no longer be interpreted as a zero-point energy.

Notice that Eq.~\eqref{eqn:Apm_p} can be inverted to give
\begin{eqnarray}
\label{AppAm}
    A^+ + A^- &\hspace{-0.2cm} =\hspace{-0.2cm}  & \omega \sqrt{\frac{m}{1 + \epsilon p^2}} q,\\[2mm]
    \label{ApmAm}
    i (A^+ - A^-) \frac{1}{N-2\rho_0} &\hspace{-0.2cm} =\hspace{-0.2cm}  & \omega \hbar \epsilon \sqrt{\frac{m}{1 + \epsilon p^2}} p, 
\end{eqnarray}
where the position operator $q$ is defined as in Eq.~\eqref{Qrep} with $A=0$. 
One can then symbolically write
\begin{eqnarray}
    p &\hspace{-0.2cm} =\hspace{-0.2cm}  & \frac{1}{\sqrt{\epsilon}} \frac{i (A^+ - A^-) \frac{1}{N-2\rho_0}}{\sqrt{\omega^2 \hbar^2 \epsilon m + \left[(A^+ - A^-) \frac{1}{N-2\rho_0}\right]^2}}, \\[1mm]
    q &\hspace{-0.2cm} =\hspace{-0.2cm}  & \frac{\hbar\sqrt{\epsilon}}{\sqrt{\omega^2 \hbar^2 \epsilon m + \left[ (A^+ - A^-) \frac{1}{N-2\rho_0}\right]^2}} (A^++A^-), 
\end{eqnarray}
where we have retained the only solutions
that recover the correct behavior in the $\epsilon\rightarrow0$ limit.
Alternatively, using the properties of hyperbolic functions, the same expressions can be written in the following way
\begin{eqnarray}
    p &\hspace{-0.2cm} =\hspace{-0.2cm}  & \frac{1}{\sqrt{\epsilon}} \sinh\left\{\text{arctanh} \left[i (A^+ - A^-) \frac{1}{\omega \hbar\sqrt{\epsilon m} (N-2\rho_0)}\right]\right\},\\[2mm]
    q &\hspace{-0.2cm} =\hspace{-0.2cm}  & \frac{1}{\omega \sqrt{m}} \cosh\left\{\text{arctanh} \left[i (A^+ - A^-) \frac{1}{\omega \hbar\sqrt{\epsilon m} (N-2\rho_0)}\right]\right\} (A^+ + A^-).
\end{eqnarray}
Using Eqs.~\eqref{AppAm} and \eqref{ApmAm}, it is in fact easy to see that such expressions correspond to the position and momentum operators $q$ and $p$.

It is now possible to verify that the eigenfunctions $\psi_n$ with the norm
\begin{eqnarray}
   |\mathcal{N}_n|^2 
    &\hspace{-0.2cm}\equiv\hspace{-0.2cm}& \langle \psi_n | \psi_n \rangle
    = |\psi_0|^2 \frac{n! (-4\rho_0)^{\overline{n}}}{(1-2\rho_0)^{\overline{n}}(-2\rho_0)^{\overline{n}}}, \\[2mm]
    |\psi_0|^2
    &\hspace{-0.2cm}=\hspace{-0.2cm}& \frac{4^{- 2 \rho_0}\sqrt{\pi}\Gamma \left(\frac{1}{2} - 2 \rho_0 \right)}{\sqrt{\epsilon }\, \Gamma (1 - 2 \rho_0 )}\,, 
\end{eqnarray}
are normalized to unity with respect to the GUP-scalar product defined in Eq.~\eqref{scpr}. 
Here $\alpha^{\overline{n}}$ is the Pochhammer symbol
\begin{equation}
    (\alpha)^{\overline{n}} = \alpha (\alpha + 1) \cdots (\alpha + n - 1).
\end{equation}
Using the above result, it is convenient to redefine the normalized energy eigenfunctions with an abuse of notation as
\begin{equation}
\label{normpsi}
    \psi_n(\zeta) = \frac{(-1)^n}{\mathcal{N}_n} \zeta^{\rho_n} (1-\zeta)^{\rho_n} \, _2F_1(a_n,b_n;c_n;\zeta), 
\end{equation}
where the factor $(-1)^n$ allows us to recover the standard expression for $\psi_n(\zeta)$ in the $\varepsilon\rightarrow0$ limit.

Thus, by introducing the Dirac notation and denoting by $|n\rangle$ the normalized energy eigenstate corresponding to the $n$-th energy level, we have
\begin{eqnarray}
\label{norm1}
    A^+ |n\rangle &\hspace{-0.2cm} =\hspace{-0.2cm}  & \frac{\omega \hbar \sqrt{\epsilon m}}{2} \sqrt{\frac{(n+1) (n-2\rho_0) (n-4\rho_0)}{n+1-2\rho_0}} |n+1\rangle, \\ [1mm]
    \label{norm2}
    A^- |n\rangle &\hspace{-0.2cm} =\hspace{-0.2cm}  & \frac{\omega \hbar \sqrt{\epsilon m}}{2} \sqrt{\frac{n (n-2 \rho_0) (n-1-4\rho_0)}{n-1-2\rho_0}} |n-1\rangle.
\end{eqnarray}

At this stage, it must be noted that $A^+$ and $A^-$ are not the adjoint of each other.
In fact, it is straightforward to see that the following relations hold
\begin{equation}
    (A^\pm)^\dagger (N-2 \rho_0)
    = (N-2 \rho_0) A^\mp.
    \label{ApiudagAmeno}
\end{equation}
Therefore, the matrix elements of $(A^\pm)^\dagger$ and $A^\mp$ in the basis of normalized states $|n\rangle$ are not equal, but differ by 
\begin{equation}
  \frac{\langle n\mp1 | (A^\pm)^\dagger |n\rangle}{\langle n\mp1 | A^{\mp} |n\rangle} =  \frac{n\mp1-2\rho_0}{n-2\rho_0}\,.
\end{equation}
However, $(A^+)^\dagger$ still acts by lowering the HO state by one, while $(A^-)^\dagger$ acts by raising the state by one.
These considerations suggest that
$A^{\pm}$ are not yet eligible candidates for the r\^ole of
generalized ladder operators, although providing a good starting point for constructing them.

With the purpose of finding the proper generalized ladder operators and following Ref.~\cite{Bagarello}, it is convenient to introduce new operators $C^\pm$ such that
\begin{eqnarray}
\label{Cp}
    C^+
    &\hspace{-0.2cm} =\hspace{-0.2cm}  & \sqrt{2}\sqrt{(N-2\rho_0) (N-4\rho_0)} A^+ \sqrt{\frac{1}{(N-2\rho_0)(N-4\rho_0)}}
    ,\\[2mm]
    \label{Cm}
    C^-
   &\hspace{-0.2cm}  =\hspace{-0.2cm}  & \sqrt{2}\sqrt{\frac{N-2\rho_0}{N-4\rho_0}} A^- \sqrt{\frac{N-4\rho_0}{N-2\rho_0}}.
\end{eqnarray}
It is fairly immediate to see $C^{\pm}$ behave as full-fledged ladder operators for the GUP-harmonic oscillator.
First, from Eq.~\eqref{ApiudagAmeno} one can prove that they are each other's Hermitian conjugates, \emph{i.e.} $(C^\pm)^\dagger=C^\mp$. For this reason, from now on we are allowed to drop the $\pm$ index and use the simpler notation $C^-\equiv C$ and $C^+\equiv C^\dagger$.  
Furthermore, they inherit the raising and lowering character of the operators $A^\pm$.
In fact, we have
\begin{equation}
    C^\dagger|n\rangle
    =  \sqrt{E_{n+1}-E_{0}}\,|n+1\rangle,\qquad\,\,\,\,
    C|n\rangle
    =  \sqrt{E_{n}-E_{0}}\,|n-1\rangle,
\end{equation}
where the energy spectrum $E_n$ is defined as in Eq.~\eqref{Enspec}.
We also notice that  $C$ and $C^\dagger$ retain the  correct $\varepsilon\rightarrow0$ limit, since they reduce to the standard annihilators and creators for the harmonic oscillator, up to the dimensional factor $\sqrt{\hbar\omega}$, 
\begin{eqnarray}
  \lim_{\epsilon\rightarrow0} C^\dagger|n\rangle&\hspace{-2mm}=\hspace{-2mm}&\sqrt{\hbar\omega}\sqrt{n+1}|n+1\rangle,\\[2mm]
   \lim_{\epsilon\rightarrow0} C|n\rangle&\hspace{-2mm}=\hspace{-2mm}&\sqrt{\hbar\omega}\sqrt{n}|n-1\rangle\,.
\end{eqnarray}

The advantage of this new set of operators is that they allow to factorize the hamiltonian in the canonical form
\begin{equation}
\label{usform}
    H= C^\dagger C + E_{0}\mathds{1}\,.
\end{equation}
Additionally, they obey the following relations
\begin{equation}
    C\hspace{0.4mm}C^\dagger=g(H)- E_{0}\mathds{1},\qquad\,\,\,\, [C,C^\dagger]=g(H)-H, \label{eqn:C_algebra}
\end{equation}
where the characteristic function $g$ is defined recursively as  $E_{n+1}=g(E_{n})$.
In our case, it takes the form
\begin{equation}
    g(H) = H + \frac{\hbar\omega}{2}\left(2\sqrt{1+2\,\epsilon m H}+\hbar\omega \epsilon m\right). \label{eqn:g(H)}
\end{equation}
Finally, it is possible to express the position and momentum operators $q$ and $p$ in terms of the operators $C$, $C^\dagger$ by first inverting Eqs.~\eqref{Cp} and~\eqref{Cm} and then using Eqs.~\eqref{AppAm} and~\eqref{ApmAm}.
We then find
\begin{eqnarray}
\label{modp}
    p &\hspace{-0.5cm}=\hspace{-0.5cm} & -\frac{1}{\sqrt{\epsilon}}\sinh\left[\text{arctanh}\left(\frac{i}{\sqrt{2\varepsilon m}\omega \hbar} \sqrt{\frac{N-4\rho_0}{N-2\rho_0}}\, C \sqrt{\frac{1}{(N-2\rho_0)(N-4\rho_0)}} + \text{h.c.}
    \right) 
    \right],\\[2mm]
\nonumber
    q &\hspace{-0.5cm}= \hspace{-0.5cm}& 
        \frac{1}{\sqrt{2m\omega^2}}\cosh{\left[\text{arctanh}\left(\frac{i}{\sqrt{2\varepsilon m}\omega \hbar} \sqrt{\frac{N-4\rho_0}{N-2\rho_0}}\, C \sqrt{\frac{1}{(N-2\rho_0)(N-4\rho_0)}} + \text{h.c.}\right)\right]}\\[1mm]
        &&\times \frac{1}{\sqrt{N-2\rho_0}} \left(\sqrt{N-4\rho_0}\, C\, \frac{1}{\sqrt{N-4\rho_0}} + \text{h.c.}\right) \sqrt{N-2\rho_0}.
        \label{modq}
\end{eqnarray}
A comment is in order here. Apart from their relevance in the theoretical study of the GUP, we emphasize that Eqs.~\eqref{modp} and \eqref{modq} might also be of interest at phenomenological level. Indeed, in Ref.~\cite{Plenio}  tests of a minimal length have been proposed based on the computation of GUP effects on the period of the pendulum and the trajectory of
the harmonic oscillator in the  Gazeau-Klauder coherent state.
In particular, the latter calculation has been carried out by resorting to the relation between the position operator and the deformed ladder operators in the approximation of small GUP parameter. Here, we have derived such relation in its exact form. We expect that this result can prove  useful in view of future high-precision experiments aimed at searching for 
quantum gravity signatures in harmonic oscillator-like systems. This application, however, goes beyond the scope of the present analysis and will be investigated elsewhere.

\section{QFT toy model}
\label{QFT}

In this section we propose a quantum-field-theoretic toy model based on the results found above.
To facilitate the description and  highlight the main points, we propose here the quantization of a free real scalar field in $(1+1)$-dimensional flat spacetime. A similar approach can in principle be followed in higher-dimensional GUP theories and/or for other spin fields. 

Following the guidelines of the standard QFT treatment, we consider the anti-transform of the annihilation operator $C_{p}$, where the subscript $p$ labels the generic field mode. We have
\begin{equation}
    \widetilde\phi(\xi) = \frac{1}{\sqrt{8\pi\hbar}} \int_{-\pi/2\sqrt{\epsilon}}^{\pi/2\sqrt{\epsilon}} dp_0 ~ \mathcal{N}_{p_0} \left[e^{i(\frac{\xi}{\hbar}+\sqrt{\epsilon}) p_0} + e^{i(\frac{\xi}{\hbar}-\sqrt{\epsilon}) p_0}\right] C_{p},  \label{eqn:field_start}
\end{equation}
where we have written the transform in terms of the auxiliary momentum $p_0$ according to the results of Sec. \ref{sec:transforms}.
Thus, the mode $p$ is to be understood as a function of  $p_0$ as in Eq.~\eqref{pp0}.
The factor $\mathcal{N}_{p_0}$ is the normalization coefficient to be fixed.

As it stands, the object in Eq.~\eqref{eqn:field_start} is not real, nor can it be made as such. We then introduce its Hermitian conjugate and build up the following self-adjoint combination
\begin{equation}
\label{fundfield}
\phi(\xi)\equiv \widetilde\phi(\xi) + \widetilde\phi^\dagger(\xi) = \frac{1}{\sqrt{8\pi\hbar}} \int_{-\pi/2\sqrt{\epsilon}}^{\pi/2\sqrt{\epsilon}} dp_0 ~ \mathcal{N}_{p_0} \left[e^{i(\frac{\xi}{\hbar}+\sqrt{\epsilon}) p_0} + e^{i(\frac{\xi}{\hbar}-\sqrt{\epsilon}) p_0}\right]  \left( C_{p} + C^{\dagger}_{-p} \right),
\end{equation}
where we have exploited the fact that $p$ is an odd function of $p_0$ in our GUP model, while $\mathcal{N}_{p_0}$ is expected to depend on the modulus of $p_0$ as in the ordinary formalism. 
The above relation provides us with the fundamental object on which our next analysis is developed\footnote{In the case of a complex field, the reasoning leading to Eq.~\eqref{fundfield} must be generalized by considering $\widetilde\phi$ and $\widetilde\phi^\dagger$ as two distinct objects. In turn, the field expansion turns out to be modified by acquiring the antiparticle degrees of freedom.}. 
In passing, we mention that a similar application of the GUP to the QFT framework has been studied in~\cite{HusQFT}. Even in that case the approach involves applying a ($3$-dimensional) spatial Fourier transform to the $\textbf{k}$-space variables and then enforcing the deformed commutator
on such variables. Corrections of the deformed commutator are then explored on first quantized-like equations of a field theory~\cite{HusQFT}. On the other hand, the r\^ole of the GUP in gauge field theories has been addressed in~\cite{Kober}, showing that the implementation of the minimal length  condition induces a generalized interaction between the matter field and the gauge field, as well as an additional self interaction of the gauge field.

Now, inspired by Eq.~\eqref{eqn:C_algebra}, we impose the following commutation relations
\begin{equation}
    [C_{p_1}, C^\dagger_{p_2}] = \alpha_p [g(\mathcal{H}_{p_1}) - \mathcal{H}_{p_1}] \delta(p_1 - p_2),
\end{equation}
with all other commutators vanishing, 
where $\mathcal{H}_p = C^\dagger_p C_p$ corresponds to the Hamiltonian density for the $p$-mode up to the zero-point energy.
Accordingly, the total Hamiltonian describing the system is
\begin{equation}
    H = \int \frac{dp}{1 + \epsilon p^2} \mathcal{H}_p.
\end{equation}
Here, $\alpha_p$ is a c-number with dimensions of energy that tends to $\hbar \omega_p$ for $\epsilon\to0$, with $\omega_p=\sqrt{m^2+p^2}/\hbar$ being the standard frequency for the $p$-mode.
Furthermore, here we will use
\begin{equation}
    g(\mathcal{H}_p) - \mathcal{H}_p
    = \sqrt{1 + 2 \epsilon m \hbar \omega_p \mathcal{H}_p} + \frac{\hbar \omega_p}{2} \epsilon m.
    \label{eqn:g_QFT}
\end{equation}
Additionally, we 
have assumed that field modes with different $p$
are still decoupled from each other.

Consistently with the canonical field quantization, we can now provide the physical interpretation of the state $\phi(\xi)|0\rangle$, where $|0\rangle$ denotes as usual the field vacuum such that
\begin{equation}
C_p|0\rangle = 0\,,\,\, \forall p.
\end{equation}
From the expansion \eqref{fundfield}, it is straightforward to check that
\begin{equation}
    \phi(\xi)|0\rangle
    = \frac{1}{\sqrt{8\pi\hbar}} \int_{-\pi/2\sqrt{\epsilon}}^{\pi/2\sqrt{\epsilon}} dp_0 ~ \mathcal{N}_{p_0} \left[e^{i(\frac{\xi}{\hbar}+\sqrt{\epsilon}) p_0} + e^{i(\frac{\xi}{\hbar}-\sqrt{\epsilon}) p_0}\right] 
    C^{\dagger}_{-p} |0\rangle.
%
\end{equation}
The above relation corresponds to a linear  superposition of single-particle states of well-definite momentum $p=p(p_0)$. 
Furthermore, it is worth noticing that
\begin{eqnarray}
\nonumber
    \langle 0 | \phi(\xi) | p \rangle
    &\hspace{-0.2cm}=\hspace{-0.2cm}& \frac{1}{\sqrt{8\pi\hbar}} \int_{-\pi/2\sqrt{\epsilon}}^{\pi/2\sqrt{\epsilon}} dp'_0 ~ \mathcal{N}_{p'_0} \left[e^{i(\frac{\xi}{\hbar}+\sqrt{\epsilon}) p'_0} + e^{i(\frac{\xi}{\hbar}-\sqrt{\epsilon}) p'_0}\right] \langle 0 | C_{p'} | p \rangle\\[2mm]
    &\sim& \frac{\mathcal{N}'_{p_0}}{\sqrt{8\pi\hbar}} \left[e^{i(\frac{\xi}{\hbar}+\sqrt{\epsilon}) p_0} + e^{i(\frac{\xi}{\hbar}-\sqrt{\epsilon}) p_0}\right]
    = \frac{\mathcal{N}'_{p_0}}{\sqrt{2\pi\hbar}} \cos (\sqrt{\epsilon} p_0) e^{i \frac{\xi}{\hbar} p_0}.
\end{eqnarray}
We thus obtain a plane wave of wavenumber $p_0/\hbar$, multiplied by a cosine factor encoding GUP corrections.
Given the similarity with the quantum mechanical case~\cite{KMM,Bosso:2020aqm}, we can interpret this as the position-space representation of the single-particle wavefunction of the state $|p\rangle$ in the GUP framework. The pure plane wave behavior is naturally recovered for $\epsilon\rightarrow0$.

It is now convenient to describe the system in the Heisenberg picture.
We then introduce the time-dependent annihilation operators
\begin{equation}
    C_p (t) \equiv e^{iHt/\hbar} C_p e^{-iHt/\hbar} = e^{-i\Omega_p t } C_p\,,
    \label{eqn:time_annihilation}
\end{equation}
and similarly for $C^\dagger_p(t)=C^\dagger_p e^{i\Omega_p t }\,$, where we have used the commutators \eqref{comQFT}
and introduced the shorthand notation
$\Omega_p \equiv [g(\mathcal{H}_p) - \mathcal{H}_p] \alpha_p/\hbar$.
Notice that the above formula gives back the standard time-dependence of the annihilation operators in the limit of vanishing GUP, since in this case $\Omega_p\rightarrow\omega_p\hspace{0.2mm}\mathbb{I}$ (see Eq.~\eqref{eqn:g_QFT}).
Thus, a remarkable property featuring the GUP framework is that the proportionality factor 
ruling the time-evolution
is now a non-trivial operator.

In what follows, we will assume that the Hamiltonian density for our free field still depends on the absolute value of the physical momentum $p$ only, \emph{i.e.} $\mathcal{H}_p\equiv\mathcal{H}(|p|)$. This is reasonable in the presence of a quadratic GUP. Combining with  Eqs.~\eqref{fundfield} and \eqref{eqn:time_annihilation}, the time-dependent field can be written as
\begin{equation}
\label{eqn:expansion}
    \phi(\xi,t) = \frac{1}{\sqrt{8\pi\hbar}} \int_{-\pi/2\sqrt{\epsilon}}^{\pi/2\sqrt{\epsilon}} dp_0 ~ \mathcal{N}_{p_0} \left[e^{i(\frac{\xi}{\hbar}+\sqrt{\epsilon}) p_0} + e^{i(\frac{\xi}{\hbar}-\sqrt{\epsilon}) p_0}\right]  \left[e^{-i\Omega_{p} t} C_{p} + C^{\dagger}_{-p} e^{+i\Omega_p t}\right].
\end{equation}
We notice that the expression for the time-dependent field reduces to the usual expression of standard quantum field theory in the limit $\epsilon \to 0$.

From the above relation, we can also compute the conjugate momentum field $\pi_0(\xi,t)$ defined as $\pi_0(\xi,t)=\frac{i}{\hbar}\left[H,\phi\right]=\dot{\phi}(\xi,t)$
\begin{equation}
    \pi_0 (\xi,t)
    = \frac{-i}{\sqrt{8\pi\hbar}} \int_{-\pi/2\sqrt{\epsilon}}^{\pi/2\sqrt{\epsilon}} dp_0 ~ \mathcal{N}_{p_0} \left[e^{i(\frac{\xi}{\hbar}+\sqrt{\epsilon}) p_0} + e^{i(\frac{\xi}{\hbar}-\sqrt{\epsilon}) p_0}\right]  \left[\Omega_p e^{-i\Omega_{p} t} C_{p} - C^{\dagger}_{-p} \Omega_p e^{+i\Omega_p t}\right].
\end{equation}
This allows us to derive the equation of motion
\begin{equation}
    \dot{\pi}_0 (\xi,t) = \ddot{\phi}(\xi,t)
    = - \frac{1}{\sqrt{8\pi\hbar}} \int_{-\pi/2\sqrt{\epsilon}}^{\pi/2\sqrt{\epsilon}} dp_0 ~ \mathcal{N}_{p_0} \left[e^{i(\frac{\xi}{\hbar}+\sqrt{\epsilon}) p_0} + e^{i(\frac{\xi}{\hbar}-\sqrt{\epsilon}) p_0}\right]  \left[\Omega_p^2 e^{-i\Omega_{p} t} C_{p} + C^\dagger_{-p} \Omega_p^2 e^{+i\Omega_p t}\right].
\end{equation}
It is convenient considering the expressions above for each mode $p$.
In this case, the field reads, as found above,
\begin{equation}
    \phi(p,t) = \mathcal{N}_p (e^{-i\Omega_p t} C_p + C^\dagger_{-p} e^{+i\Omega_p t}).
\end{equation}
Furthermore, we find
\begin{equation}
    \pi_0(p,t) = - i \mathcal{N}_p (\Omega_p e^{-i\Omega_p t} C_p - C^\dagger_{-p} \Omega_p e^{+i\Omega_p t}),
\end{equation}
and
\begin{equation}
    \dot{\pi}_0(p,t) = \ddot{\phi}(p,t) = - \mathcal{N}_p (\Omega_p^2 e^{-i\Omega_p t} C_p + C^\dagger_{-p} \Omega_p^2 e^{+i\Omega_p t}).
\end{equation}
This last expression can be written in term of the field $\phi(p,t)$ as
\begin{equation}
    \ddot{\phi}(p,t) = - \Omega_p \phi(p,t) \Omega_p + \frac{1}{2} \left[\Omega_p \phi(p,t) + \phi(p,t) \Omega_p\right] (\alpha_p \omega_p \epsilon m) - \frac{1}{2} \phi(p,t) (\alpha_p \omega_p \epsilon m)^2,
    \label{eqn:eom_p}
\end{equation}
where we used the results of Appendix \ref{Bap}.
It is worth noticing that the last two terms on the right hand side vanish in the limit $\epsilon \to 0$, thus leaving the first term and the usual equation of motion in the low energy limit.
Furthermore, since $\Omega_p$ is related to the Hamiltonian density, the presence of an interaction in the form a minimal coupling intervenes in such object.
We also observe that the equation of motion above has been found in momentum space.
In fact, due to the modification introduced here related to a minimal length, the equation of motion in configuration space becomes non-trivial, involving an integro-differential equation.
In fact, such equation of motion can be obtained by transforming Eq.~\eqref{eqn:eom_p} according to the prescription of Sec. \ref{sec:transforms}.

\section{Conclusions and Outlook}
\label{Conc}
Several models of quantum gravity 
predict the emergence of a minimal length at Planck scale.
A widespread way of dealing with this supposed quantum feature of spacetime is to modify the Heisenberg Uncertainty Principle so as to accommodate a minimal position uncertainty of order $\ell_p$. The ensuing uncertainty relation is referred to as Generalized Uncertainty Principle.
In this paper, we studied the implications of a GUP with linear and quadratic momentum corrections on the harmonic oscillator. Following the standard treatment in low-energy quantum mechanics, we presented both the analytic and algebraic analyses. Remarkably, the latter method works efficiently only with the quadratic GUP, the linear correction impeding to define ladder operators. We found that the HO eigenfunctions and spectrum are non-trivially modified, the results being consistent with previous achievements in literature. The above formalism was finally  exploited to develop a field theoretical toy model based on the GUP. 

While timely and largely investigated in literature~\cite{Pedram:2012ui,Bosso:2018syo,Dossa:2021tiq}, to the best of our knowledge this is the first time
the derivation of the generalized HO algebra is carried out exactly in terms of the position and momentum operators. For comparison, we remark that the analysis of Ref.~\cite{Pedram:2012ui} is based on the generalized Heisenberg algebra given in~\cite{Curado}, which  exploits the formal action of the 
ladder operators on the Fock space without referring to their explicit dependence on $p$ and $q$. 
On other hand, in~\cite{Bosso:2018syo} it was proposed a perturbative construction of the HO algebra for the case of an arbitrary self-adjoint polynomial perturbation of the commutator. Apart from its intrinsic theoretical interest, our study might be relevant for experimental purposes as well. 
For instance, in~\cite{Plenio} 
phenomenological tests of the GUP
were explored in connection with optomechanical system and oscillator frequency measurements. In particular,  concerning the latter framework the strategy is to search for quantum gravity signatures by looking at deviations of the oscillatory trajectory from the expected result. In Ref.~\cite{Plenio} this was done by computing the expectation
value of the HO position in the Gazeau-Klauder state through the  relation between $q$ and $a$, $a^\dagger$ approximated for small  GUP parameter. 
Although quite satisfactory in 
describing QG effects at present, it is reasonable to expect that quantum tests of gravity might achieve an increasingly high level of precision in the next future, thus demanding theoretical models to be as accurate as possible. From this perspective, our work is to be intended as a further step towards the full theoretical understanding of GUP effects on harmonic oscillator-like systems. 

Despite the non-trivial results, some issues are yet to be investigated. For instance, in our analysis we resorted to the most common polynomial form of GUP. Nevertheless, several deformations of the HUP with a minimal length have been proposed in the last years~\cite{Nou,Pedrambis,Chung}. The question thus arises as to how our formalism and results are modified in these alternative frameworks. In particular, it would be interesting to investigate the impact of different commutation rules on the field quantization. Along this line, a more demanding task is to explore the consequences of a minimum length in a fully covariant Quantum Field Theory. Notice that a similar problem has been recently addressed in
Refs.~\cite{BossoDas:2020}, where a covariant generalization of the GUP with a Lorentz invariant minimum length~\cite{BosTodo} was used to analyze quantum gravity effects on scattering amplitudes in scalar electrodynamics. 
Clearly, such a covariant formulation would also pave the way for a direct extension of the QFT with the GUP to some other background besides Minkowski spacetime. Of particular interest is the field quantization on black hole geometry. 

Finally, given the new definition of ladder operators with respect to the previous literature, the formalism presented in the current work allows for a new study of coherent states in both quantum mechanics and field theory. For instance, inspired by Ref.~\cite{BossoCohe}, one might 
extend the above quantization to the electromagnetic field and  compute Planck scale corrections to sources of noise in interefrometry experiments, with particular focus on gravity wave detectors such as LIGO. The strategy is to evaluate the differential momentum transferred to the end
mirrors of the two arms of the interferometer. This can be done
by computing the expectation value of the difference of photon number in each arm, considering a coherent state for one input channel of the interferometer and a squeezed state for the other~\cite{Caves}.  In the presence of GUP effects, one then expects a shift in the interference fringes, due to a pressure gradient experienced by the two mirrors. 
Therefore, the analysis of noise data from advanced interferometry experiments may provide 
very stringent bounds on the existence of a
minimal length, if not an indirect observation of it.
Work along the above directions is under active investigation and will be presented elsewhere.

\begin{acknowledgements}

\noindent The authors are grateful to Fabio Bagarello (Universit\`a degli Studi di Palermo) for helpful conversations. 
\end{acknowledgements}


\begin{appendices}

\setcounter{equation}{0}
\renewcommand{\theequation}{\thesection\arabic{equation}}

\section{\hspace{-0.55cm}: Functions of canonical variables in quantum mechanics}
\label{funcanvar}

Here, we will review the steps to derive a Taylor series in two dimensions, accommodating it for non-commutating variables.
In the same hypothesis as for the ordinary two-dimensional Taylor series, we consider a function $f=f(Q,P)$ of two variables such that $[Q,P]\neq0$.
Furthermore, let $x$ and $y$ be a pair of constants, representing the point about which we want to construct the series.
Given these constants, we can define
\begin{equation}
	\tilde{Q}(t) =  x + t (Q - x), \qquad\,\,\,\,
	\tilde{P}(t) =  y + t (P - y).
\end{equation}
Thus, we can introduce a function $F(t) = f(\tilde{Q}(t),\tilde{P}(t))$ such that
\begin{equation}
	F(0) =  f(x,y), \qquad\,\,\,\,
	F(1) =  f(Q,P).
\end{equation}
By expanding $F(t)$ in series around $t=0$, we have
\begin{equation}
	F(t) = \sum_{n=0}^\infty \frac{t^n}{n!} \left. \frac{\diff^n F}{\diff t^n} \right|_{t=0}
	= \sum_{n=0}^\infty \frac{t^n}{n!} \sum_{a=0}^n \left. \frac{\partial^n f}{\partial Q^a \partial P^{n-a}} \right|_{Q=x,P=y} \sigma(a,n-a), \label{eqn:taylor2}
\end{equation}
with $\sigma(a,b)$ being the sum of all the permutations of $a$ copies of $Q-x$ and $b$ copies of $P-y$.
Clearly, we obtain the desired series for $t=1$.
It is worth observing that $\sigma(a,b)$ is Hermitian.

In what follows, it is convenient to have $\sigma(a,b)$ as a sum of terms with all the position operators on one side and all the momentum operators on the other.
We then have the following relations
\begin{subequations}
\begin{equation}
	\sigma(a,b) =  \sum_{j=0}^{\min(a,b)} \frac{(i \hbar)^j}{(2j)!!} \frac{(a+b)!}{(a-j)!(b-j)!} (P-y)^{b-j} (Q-x)^{a-j}
	\end{equation}
	\begin{equation}
	\hspace{12.5mm} =  \sum_{j=0}^{\min(a,b)} \frac{(- i \hbar)^j}{(2j)!!} \frac{(a+b)!}{(a-j)!(b-j)!} (Q-x)^{a-j} (P-y)^{b-j}.
	\end{equation}
\end{subequations}
Specializing to the case $[Q,P]=i\hbar$ and using the following relations
\begin{equation}
    [Q^a,P^b]
    = \sum_{j=1}^{\min(a,b)} \frac{(i \hbar)^j}{j!} a^{\underline{j}}\hspace{0.5mm} b^{\underline{j}} ~ P^{b-j} Q^{a-j} 
    = - \sum_{j=1}^{\min(a,b)} \frac{(-i \hbar)^j}{j!} a^{\underline{j}}\hspace{0.5mm} b^{\underline{j}} ~ Q^{a-j} P^{b-j},
\end{equation}
we obtain
\begin{eqnarray}
\hspace{-4mm}	[\sigma(a,b),Q^A] 
	&\hspace{-0.2cm}=\hspace{-0.2cm} & - \sum_{j=1}^{\min(A,b)} \frac{(i\hbar)^j}{j!} (a+b)^{\underline{j}} ~ A^{\underline{j}} ~ \sigma(a,b-j) Q^{A-j} 
	= \sum_{j=1}^{\min(A,b)} \frac{(- i\hbar)^j}{j!} (a+b)^{\underline{j}} ~ A^{\underline{j}} ~ Q^{A-j} \sigma(a,b-j),\\[2mm]
\hspace{-4mm}	[\sigma(a,b),P^B] 
	&\hspace{-0.2cm}=\hspace{-0.2cm} & \sum_{j=1}^{\min(a,B)} \frac{(i\hbar)^j}{j!} (a+b)^{\underline{j}} ~ B^{\underline{j}} ~ P^{B-j} \sigma(a-j,b) 
	= - \sum_{j=1}^{\min(a,B)} \frac{(- i\hbar)^j}{j!} (a+b)^{\underline{j}} ~ B^{\underline{j}} ~ \sigma(a-j,b) P^{B-j},
\end{eqnarray}
where we have introduced the shorthand notation $n^{\underline{m}} = n! / (n-m)!$ to describe a falling factorial.

Using the above results, we can write
\begin{eqnarray}
	[f(q,p),Q^A] 
	&\hspace{-0.2cm}=\hspace{-0.2cm} & - \sum_{j=1}^A \frac{(i \hbar)^j}{j!} A^{\underline{j}} ~ \frac{\partial^j f}{\partial P^j} Q^{A-j} 
	\,=\, \sum_{j=1}^A \frac{(- i \hbar)^j}{j!} A^{\underline{j}} ~ Q^{A-j} \frac{\partial^j f}{\partial P^j},\\[2mm]
	[f(q,p),P^B] 
	&\hspace{-0.2cm}=\hspace{-0.2cm} & \sum_{j=1}^B \frac{(i \hbar)^j}{j!} B^{\underline{j}} ~ P^{B-j} \frac{\partial^j f}{\partial Q^j} 
	\,=\, - \sum_{j=1}^B \frac{(- i \hbar)^j}{j!} B^{\underline{j}} ~ \frac{\partial^j f}{\partial Q^j} P^{B-j}.
\end{eqnarray}
Here, differentiating the function $f(Q,P)$ with respect to $Q$ or $P$ amounts to considering $f(Q,P)$ as a c-valued function of the quantities $Q$ and $P$ and differentiating with respect to them.
Only after having performed the differentiation, we can consider $Q$, $P$ and the derivatives of the function $f(Q,P)$ as operators again.
To distinguish between the c-valued and q-valued quantities, we will mark the second by a $\widehat{\,\cdot\,}$.
In this way, the relations above become
\begin{eqnarray}
\label{comfq}
	[f(\hat{Q},\hat{P}),\hat{Q}^A] 
	&\hspace{-0.2cm}=\hspace{-0.2cm} & - \sum_{j=1}^A \frac{(i \hbar)^j}{j!} A^{\underline{j}} ~ \widehat{\frac{\partial^j f}{\partial P^j}} \hat{Q}^{A-j} 
	\,=\, \sum_{j=1}^A \frac{(- i \hbar)^j}{j!} A^{\underline{j}} ~ \hat{Q}^{A-j} \widehat{\frac{\partial^j f}{\partial P^j}},\\[2mm]
	[f(\hat{Q},\hat{P}),\hat{P}^B] 
&\hspace{-0.2cm}=\hspace{-0.2cm} & \sum_{j=1}^B \frac{(i \hbar)^j}{j!} B^{\underline{j}} ~ \hat{P}^{B-j} \widehat{\frac{\partial^j f}{\partial Q^j}} 
	\,=\, - \sum_{j=1}^B \frac{(- i \hbar)^j}{j!} B^{\underline{j}} ~ \widehat{\frac{\partial^j f}{\partial Q^j}} \hat{P}^{B-j}.
	\label{comfp}
\end{eqnarray}
Furthermore, we have the following relations
\begin{eqnarray}
\nonumber
 \hspace{-1cm}   [\sigma(a,b),\sigma(A,B)] &\hspace{-0.2cm}=\hspace{-0.2cm} & 
         \sigma(a,b) \sigma(A,B) \\[1.5mm]
        && - \sum_{j=0}^{\min(a,B)} \sum_{j'=0}^{\min(A,b)} \frac{(-i\hbar)^j}{j!} \frac{(i\hbar)^{j'}}{j'!} (a+b)^{\underline{j+j'}}\hspace{0.5mm} (A+B)^{\underline{j+j'}}\hspace{0.5mm} \sigma(a-j,b-j') \sigma(A-j',B-j)
    \\[1.5mm]
\nonumber
    &\hspace{-0.2cm}=\hspace{-0.2cm} & 
         \sum_{j=0}^{\min(a,B)} \sum_{j'=0}^{\min(A,b)} \frac{(i\hbar)^{j}}{j!} \frac{(-i\hbar)^{j'}}{j'!} (a+b)^{\underline{j+j'}}\hspace{0.5mm} (A+B)^{\underline{j+j'}}\hspace{0.5mm} \sigma(A-j',B-j) \sigma(a-j,b-j') \\[1.5mm]
        && - \sigma(A,B) \sigma(a,b),
\end{eqnarray}
from which
\begin{multline}
    [\sigma(a,b),\sigma(A,B)] = \sum_{j=0}^{\min(a,B)} \sum_{j'=0}^{\min(A,b)} \frac{(i\hbar)^{j+j'}}{j! j'!} (a+b)^{\underline{j+j'}}\hspace{0.5mm} (A+B)^{\underline{j+j'}}\\[1.5mm]
    \times \left[(-1)^{j'} \sigma(A-j',B-j) \sigma(a-j,b-j') - (-1)^j \sigma(a-j,b-j') \sigma(A-j',B-j)\right].
\end{multline}

The result above allows us to find the commutator between two functions $f=f(\hat{Q},\hat{P})$ and $g=g(\hat{Q},\hat{P})$, \emph{i.e.}
\begin{eqnarray}
\nonumber
	[f(\hat{Q},\hat{P}),g(\hat{Q},\hat{P})]
	&\hspace{-0.2cm}=\hspace{-0.2cm}& \frac{1}{2} \sum_{\substack{j,j'=0\\j+j'>0}}^\infty \frac{(i \hbar)^{j+j'}}{j! j'!} \left[(-1)^j \widehat{\frac{\partial^{j+j'} g}{\partial Q^j \partial P^{j'}}} \widehat{\frac{\partial^{j+j'} f}{\partial Q^{j'} \partial P^j}} - (-1)^{j'} \widehat{\frac{\partial^{j+j'} f}{\partial Q^{j'} \partial P^j}} \widehat{\frac{\partial^{j+j'} g}{\partial Q^j \partial P^{j'}}}\right]\\[2mm]
   &\hspace{-0.2cm} =\hspace{-0.2cm}& \frac{1}{2} \sum_{\substack{j,j'=0\\j+j'>0}}^\infty \frac{(i \hbar)^j (- i \hbar)^{j'}}{j! j'!} \left[\widehat{\frac{\partial^{j+j'} g}{\partial Q^{j'} \partial P^j}} \widehat{\frac{\partial^{j+j'} f}{\partial Q^j \partial P^{j'}}} - \widehat{\frac{\partial^{j+j'} f}{\partial Q^{j'} \partial P^j}} \widehat{\frac{\partial^{j+j'} g}{\partial Q^j \partial P^{j'}}}\right].
\end{eqnarray}
It is worth noticing that in general $[\hat{f},\hat{g}] \neq i \hbar \widehat{\{f,g\}}$, with $\{\cdot,\cdot\}$ representing the Poisson brackets of classical mechanics.

As a particular example, consider the case in which $f$ and $g$ are functions of only one variable, that is $f=f(\hat Q)$ and $g=g (\hat P)$.
In this case, we have
\begin{equation}
	[f(\hat{Q}),g(\hat{P})] = \frac{1}{2} \sum_{j=1}^\infty \frac{(i \hbar)^{j}}{j!} \left[\widehat{\frac{\partial^{j} g}{\partial P^{j}}} \widehat{\frac{\partial^{j} f}{\partial Q^{j}}} - (-1)^{j} \widehat{\frac{\partial^{j} f}{\partial Q^{j}}} \widehat{\frac{\partial^{j} g}{\partial P^{j}}}\right]. \label{eqn:com_ex}
\end{equation}
We then see that the term with $j=1$ corresponds to the Poisson bracket $\{f,g\}$ once a symmetric ordering prescription has been applied.
The other terms, \emph{i.e.} for $j>1$, cannot be accounted for in the same way.

For the sake of concreteness, we consider the case $f(\hat Q)=\hat Q^4$, $g(\hat P)=\hat P^3$.
We then have
\begin{equation}
	[\hat{Q}^4,\hat{P}^3] =  6 i \hbar \left(\hat{P}^2 \hat{Q}^3 + \hat{Q}^3 \hat{P}^2\right) - 12 (i\hbar)^3 \hat{Q}, \qquad\,\,\,\,
	\widehat{\{Q^4,P^3\}} = 6 \left(\hat{P}^2 \hat{Q}^3 + \hat{Q}^3 \hat{P}^2\right).
\end{equation}
Suppose now that we want to change the variables from $Q$ and $P$ to $\tilde{Q}=\tilde{Q}(Q,P)$ and $\tilde{P}=\tilde{P}(Q,P)$ such that
\begin{equation}
	[\hat{Q},\hat{P}]=[\hat{\tilde{Q}},\hat{\tilde{P}}] = i\hbar.
	\label{commutator}
\end{equation}
According to the results above, we need to have
\begin{equation}
	[\tilde{Q}(\hat{Q},\hat{P}),\tilde{P}(\hat{Q},\hat{P})] = \frac{1}{2} \sum_{\substack{j,j'=0\\j+j'>0}}^\infty \frac{(i \hbar)^j (- i \hbar)^{j'}}{j! j'!} \left[\reallywidehat{\frac{\partial^{j+j'} \tilde{P}}{\partial Q^{j'} \partial P^j}} \reallywidehat{\frac{\partial^{j+j'} \tilde{Q}}{\partial Q^j \partial P^{j'}}} - \reallywidehat{\frac{\partial^{j+j'} \tilde{Q}}{\partial Q^{j'} \partial P^j}} \reallywidehat{\frac{\partial^{j+j'} \tilde{P}}{\partial Q^j \partial P^{j'}}}\right] = i \hbar. \label{eqn:com_canonical}
\end{equation}
For this to happen, it is sufficient requiring that
\begin{equation}
	[\hat{\tilde{Q}},\hat{\tilde{P}}] = i \hbar \reallywidehat{\{\tilde{Q},\tilde{P}\}} \qquad \text{with} \quad \{\tilde{Q},\tilde{P}\} = 1.
\end{equation}
In fact, when the last condition is satisfied, $\tilde{Q}$ and $\tilde{P}$ form a set of canonical variables.
Since the same is true for $Q$ and $P$, we can define a canonical transformation between the two sets.
In particular we find
\begin{equation}
	\frac{\partial \tilde{P}}{\partial Q} =  - \frac{\partial P}{\partial \tilde{Q}}, \qquad\,\,\,\,
	\frac{\partial \tilde{Q}}{\partial Q} =  \frac{\partial P}{\partial \tilde{P}}, \qquad\,\,\,\,
	\frac{\partial \tilde{P}}{\partial P} =  \frac{\partial Q}{\partial \tilde{Q}}, \qquad\,\,\,\,
	\frac{\partial \tilde{Q}}{\partial P} =  - \frac{\partial Q}{\partial \tilde{P}}. \label{eqn:derivatives}
\end{equation}
We then see that any term in Eq.~\eqref{eqn:com_canonical} with $j+j'>1$ vanishes.
Furthermore, by the same Eq.~\eqref{eqn:derivatives}, we can find that all these derivatives, considered as quantum operators, commute with each other.
Therefore, we can safely write
\begin{equation}
	[\tilde{Q}(\hat{Q},\hat{P}),\tilde{P}(\hat{Q},\hat{P})] = i \hbar \left(\reallywidehat{\frac{\partial \tilde{Q}}{\partial Q}} \reallywidehat{\frac{\partial \tilde{P}}{\partial P}} - \reallywidehat{\frac{\partial \tilde{P}}{\partial Q}} \reallywidehat{\frac{\partial \tilde{Q}}{\partial P}} \right) = i \hbar \reallywidehat{\{\tilde{Q},\tilde{P}\}}.
\end{equation}
Using similar arguments, it is possible to show that the commutator of two generic functions does not depend on the particular set of canonical coordinates, that is
\begin{equation}
	[f(Q,P),g(Q,P)] = [f(Q(\tilde{Q},\tilde{P}),P(\tilde{Q},\tilde{P})),g(Q(\tilde{Q},\tilde{P}),P(\tilde{Q},\tilde{P}))].
\end{equation}
On the other hand, allowing any of the other terms in Eq.~\eqref{eqn:com_canonical} not to vanish would imply $[\hat{\tilde{Q}},\hat{\tilde{P}}] \neq i \hbar \reallywidehat{\{\tilde{Q},\tilde{P}\}}$ and the arguments above do not apply.
In particular, we cannot be assured that the commutators remain the same in the two sets of variables.

\section{\hspace{-0.55cm}: Commutators involving ladder operators}
\label{Bap}
Let us start from the commutator between the $C$-operators in Eq.~\eqref{eqn:C_algebra}
\begin{equation}
    [C,C^\dagger] = g(H)-H.
\end{equation}
Then, given the form of the Hamiltonian in terms of the $C$-operators in Eq.~\eqref{usform}, we have
\begin{equation}
\label{CmenoHcom}
    [C,H]=[g(H)-H] C.
\end{equation}
Furthermore, by induction we can find
\begin{equation}
    [C,H^n] = \sum_{i=1}^n \binom{n}{i} [g(H)-H]^i H^{n-i} C
    = [g^n(H) - H^n] C.
\end{equation}

The result above allows for finding the commutator of $C$ and $f(H)$, with $f(x)$ an infinitely differentiable function at $x=0$.
In fact, expanding in Taylor series the function $f(H)$ in the following commutator, we get
\begin{equation}
    [C, f(H)]
    = \sum_{j=0}^\infty \frac{1}{j!} \left. \frac{d^j f(x)}{d x^j}\right|_{x=0} [g^j(H) - H^j] C
    = [f(g(H)) - f(H)] C.
\end{equation}
Clearly, the same result can be applied to the case of the commutator of $C$ and $g(H)$, for which we find
\begin{equation}
    [C, g(H)]
    = [g(g(H)) - g(H)] C.
\end{equation}
Furthermore, we have
\begin{eqnarray}
\nonumber
    [C,f(g^{-1}(H))]
    &\hspace{-0.2cm}=\hspace{-0.2cm}& [f(H) - f(g^{-1}(H))] C
    = C f(g^{-1}(H)) - f(g^{-1}(H)) C\\[2mm]
    &\hspace{-0.2cm}=\hspace{-0.2cm}& C[f(H) - f(g^{-1}(H))] - [f(g(H)) - f(H)] C + [C,f(g^{-1}(H))].
\end{eqnarray}
From which, we obtain
\begin{equation}
    C[f(H) - f(g^{-1}(H))] = [f(g(H)) - f(H)] C,
\end{equation}
and
\begin{equation}
    [C,f(H)]
    = C[f(H)-f(g^{-1}(H))]
    \qquad \Rightarrow \qquad
    f(H) C = C f(g^{-1}(H)).
\end{equation}
Similar relations are valid for $C^\dagger$.
Specifically,
\begin{equation}
    [C^\dagger,f(H)]
    = - C^\dagger [f(g(H)) - f(H)] = - [f(H) - f(g^{-1}(H))] C^\dagger,
\end{equation}
and
\begin{equation}
    C^\dagger f(H) = f(g^{-1}(H)) C^\dagger.
\end{equation}

We now consider the commutation relations between the ladder operators and exponential functions of the Hamiltonian.
Based on the results above, we find
\begin{eqnarray}
\label{ComCmexp}
    [C,e^{iHt/\hbar}]
    &\hspace{-0.2cm}=\hspace{-0.2cm}& [e^{ig(H)t/\hbar} - e^{iHt/\hbar}] C = C[e^{iHt/\hbar} - e^{ig^{-1}(H)t/\hbar}],\\[2mm]
\label{ComCpexp}
    [C^\dagger,e^{iHt/\hbar}]
    &\hspace{-0.2cm}=\hspace{-0.2cm}& - [e^{iHt/\hbar} - e^{ig^{-1}(H)t/\hbar}] C^\dagger = - C^\dagger [e^{ig(H)t/\hbar} - e^{iHt/\hbar}].
\end{eqnarray}

\subsection{\hspace{-0.46cm}: QFT extension} 
\label{sapx:QFT}

In the case of QFT, we have the following commutators
\begin{equation}
    [C_p,H] = \int_{-\pi/2\sqrt{\epsilon}}^{\pi/2\sqrt{\epsilon}} dp'_0 [C_p,\mathcal{H}_{p'}]
    = \int_{-\pi/2\sqrt{\epsilon}}^{\pi/2\sqrt{\epsilon}} dp'_0\, \hbar \Omega_p\, C_p\, \delta(p_0,p'_p)
    = \hbar\, \Omega_p C_p,
\end{equation}
with
\begin{equation}
    \Omega_p = [g(\mathcal{H}_p) - \mathcal{H}_p] \frac{\alpha_p}{\hbar},
\end{equation}
and $\alpha_p$ defined as in Sec.~\ref{QFT}.

It is easy to see that the relations found in QM are valid here as well when a function of the Hamiltonian density is concerned rather than the full Hamiltonian, namely
\begin{equation}
\label{comQFT}
    f(\mathcal{H}_p) C_p =  C_p f(g^{-1}(\mathcal{H}_p)),\qquad\,\,\,\, 
    C^\dagger_p f(\mathcal{H}_p) =  f(g^{-1}(\mathcal{H}_p)) C^\dagger_p.
\end{equation}
In the explicit case presented here, in which
\begin{equation}
    g(\mathcal{H}_p) - \mathcal{H}_p = \sqrt{1 + 2 \epsilon m \hbar \omega_p \mathcal{H}_p} + \frac{\hbar \omega_p}{2} \epsilon m,
\end{equation}
we find
\begin{equation}
    g^{-1}(\mathcal{H}_p) - \mathcal{H}_p = - \sqrt{1 + 2 \epsilon m \hbar \omega_p \mathcal{H}_p} + \frac{\hbar \omega_p}{2} \epsilon m
    = - [g(\mathcal{H}_p) - \mathcal{H}_p] + \hbar \omega_p \epsilon m
    = - \Omega_p \frac{\hbar}{\alpha_p} + \hbar \omega_p \epsilon m.
\end{equation}

\end{appendices}


\begin{thebibliography}{99}

\bibitem{QG1}
 D. Amati, M. Ciafaloni and G. Veneziano, Phys. Lett. B \textbf{197}, 81 (1987).
 
 \bibitem{QG2}
D. J. Gross and P. F. Mende, Phys. Lett. B 197, \textbf{129} (1987).

\bibitem{QG3}
D. Amati, M. Ciafaloni and G. Veneziano, Phys. Lett. B \textbf{216}, 41 (1989).

\bibitem{QG4}
C.~Rovelli and L.~Smolin,
Nucl. Phys. B \textbf{331}, 80 (1990).

\bibitem{QG5}
K. Konishi, G. Paffuti and P. Provero, Phys. Lett. B \textbf{234}, 276 (1990).

\bibitem{QG6}
 M. Maggiore, Phys. Lett. B \textbf{304}, 65 (1993); Phys. Lett. B \textbf{319}, 83 (1993). 
 
 
\bibitem{AdSan}
R. J. Adler and D. I. Santiago, Mod. Phys. Lett. A \textbf{14}, 1371 (1999).

\bibitem{QG8}
S. Capozziello, G. Lambiase and G. Scarpetta, Int. J. Theor. Phys. \textbf{39}, 15 (2000).

\bibitem{QG9}
J.~Magueijo and L.~Smolin,
Phys. Rev. D \textbf{71}, 026010 (2005).

\bibitem{QG10}
G.~Amelino-Camelia,
Symmetry \textbf{2}, 230-271 (2010).

\bibitem{ScardBH}
F. Scardigli, Phys. Lett. B \textbf{452}, 39 (1999).

\bibitem{KMM}
A. Kempf, G. Mangano and R. B. Mann, Phys. Rev. D \textbf{52}, 1108 (1995).

\bibitem{Bosso:2020aqm}
P.~Bosso,
Class. Quant. Grav. \textbf{38}, 075021 (2021).

\bibitem{DasPrl}
S. Das and E. C. Vagenas, Phys. Rev. Lett. \textbf{101}, 1 (2008).

\bibitem{Jizba}
P.~Jizba, H.~Kleinert and F.~Scardigli,
Phys. Rev. D \textbf{81}, 084030 (2010).


\bibitem{FrPan}
A.~M.~Frassino and O.~Panella,
Phys. Rev. D \textbf{85}, 045030 (2012). 

\bibitem{HusQFT}
V.~Husain, D.~Kothawala and S.~S.~Seahra,
Phys. Rev. D \textbf{87}, 025014 (2013).

\bibitem{ScardLamb}
F.~Scardigli, G.~Lambiase and E.~Vagenas,
Phys. Lett. B \textbf{767}, 242 (2017).

\bibitem{BossoCohe}
P.~Bosso, S.~Das and R.~B.~Mann,
Phys. Rev. D \textbf{96},  066008 (2017).

\bibitem{ScardLuc}
F.~Scardigli, M.~Blasone, G.~Luciano and R.~Casadio,
Eur. Phys. J. C \textbf{78}, 728 (2018).

\bibitem{Lake:2018zeg}
M.~J.~Lake, M.~Miller, R.~F.~Ganardi, Z.~Liu, S.~D.~Liang and T.~Paterek,
Class. Quant. Grav. \textbf{36}, 155012 (2019).


\bibitem{LucPetr}
G.~G.~Luciano and L.~Petruzziello,
Eur. Phys. J. C \textbf{79}, 283 (2019). 

\bibitem{BossoDas:2020}
P.~Bosso, S.~Das and V.~Todorinov,
Annals Phys. \textbf{422}, 168319 (2020);
Annals Phys. \textbf{424}, 168350 (2021). 

\bibitem{Blasone:2019wad}
M.~Blasone, G.~Lambiase, G.~G.~Luciano, L.~Petruzziello and F.~Scardigli,
Int. J. Mod. Phys. D \textbf{29}, 2050011 (2020).

\bibitem{Shababi}
H.~Shababi and K.~Ourabah,
Eur. Phys. J. Plus \textbf{135}, 697 (2020).

\bibitem{ShabaLuc}
G.~G.~Luciano,  Eur. Phys. J. C \textbf{81}, 672 (2021).


\bibitem{Luciano:2021cna}
G.~G.~Luciano and L.~Petruzziello,
Eur. Phys. J. Plus \textbf{136}, 179 (2021).



\bibitem{BH1}
R.~J.~Adler, P.~Chen and D.~I.~Santiago,
Gen. Rel. Grav. \textbf{33}, 2101 (2001).

\bibitem{BH2}
G.~Amelino-Camelia, M.~Arzano, Y.~Ling and G.~Mandanici,
Class. Quant. Grav. \textbf{23}, 2585 (2006). 

\bibitem{BH3}
Y.~S.~Myung, Y.~W.~Kim and Y.~J.~Park,
Phys. Lett. B \textbf{645}, 393 (2007).

\bibitem{BH4}
A.~Bina, S.~Jalalzadeh and A.~Moslehi,
Phys. Rev. D \textbf{81}, 023528 (2010).

\bibitem{BH5}
Y.~C.~Ong,
JCAP \textbf{09}, 015 (2018).


\bibitem{BH5bis}
A.~Alonso-Serrano, M.~P.~D\k{a}browski and H.~Gohar,
Phys. Rev. D \textbf{97}, 044029 (2018).

\bibitem{BH6}
L.~Buoninfante, G.~G.~Luciano and L.~Petruzziello,
Eur. Phys. J. C \textbf{79}, 663 (2019).

\bibitem{BH6bis}
H.~Hassanabadi, E.~Maghsoodi and W.~S.~Chung,
Eur. Phys. J. C \textbf{79}, 358 (2019). 

\bibitem{BH7}
P.~Bosso and O.~Obreg\'on,
Class. Quant. Grav. \textbf{37}, 045003 (2020).

\bibitem{BH8}
L.~Buoninfante, G.~Lambiase, G.~G.~Luciano and L.~Petruzziello,
Eur. Phys. J. C \textbf{80}, 853 (2020).

\bibitem{BH9}
D.~Chemisana, J.~Gin\'e and J.~Madrid,
EPL \textbf{130}, 60002 (2020).

\bibitem{BH10}
L.~Buoninfante, G.~G.~Luciano, L.~Petruzziello and F.~Scardigli,
[arXiv:2009.12530 [hep-th]].

\bibitem{BH11}
B.~Hamil and B.~C.~L\"utf\"uo\u{g}lu,
EPL \textbf{134}, 50007 (2021).

\bibitem{BH12}
L.~Petruzziello,
Class. Quant. Grav. \textbf{38}, 135005 (2021). 


\bibitem{GUPcosm1}
B.~Vakili,
Phys. Rev. D \textbf{77}, 044023 (2008).

\bibitem{GUPcosm2}
A.~Paliathanasis, S.~Pan and S.~Pramanik,
Class. Quant. Grav. \textbf{32}, 245006 (2015).

 
 \bibitem{Graphene}
A.~Iorio, P.~Pais, I.~A.~Elmashad, A.~F.~Ali, M.~Faizal and L.~I.~Abou-Salem,
Int. J. Mod. Phys. D \textbf{27}, 1850080 (2018).

 \bibitem{Bruk}
 I.~Pikovski, M.~R.~Vanner, M.~Aspelmeyer, M.~S.~Kim and C.~Brukner,
Nature Phys. \textbf{8}, 393 (2012).

\bibitem{AliTest}
A.~F.~Ali, S.~Das and E.~C.~Vagenas,
Phys. Rev. D \textbf{84}, 044013 (2011).

\bibitem{Bawaj}
M.~Bawaj, C.~Biancofiore, M.~Bonaldi, F.~Bonfigli, A.~Borrielli, G.~Di Giuseppe, L.~Marconi, F.~Marino, R.~Natali and A.~Pontin, \textit{et al.}
Nature Commun. \textbf{6}, 7503 (2015). 

\bibitem{Pendu}
P.~A.~Bushev, J.~Bourhill, M.~Goryachev, N.~Kukharchyk, E.~Ivanov, S.~Galliou, M.~E.~Tobar and S.~Danilishin,
Phys. Rev. D \textbf{100}, 066020 (2019).
 
 \bibitem{GravBar}
F.~Marin, F.~Marino, M.~Bonaldi, M.~Cerdonio, L.~Conti, P.~Falferi, R.~Mezzena, A.~Ortolan, G.~A.~Prodi and L.~Taffarello, \textit{et al.}
Nature Phys. \textbf{9}, 71 (2013).
 
 
 \bibitem{ScarCas}
F.~Scardigli and R.~Casadio,
Eur. Phys. J. C \textbf{75}, 425 (2015). 

\bibitem{BossoLigo}
P.~Bosso, S.~Das and R.~B.~Mann,
Phys. Lett. B \textbf{785},  498 (2018).

\bibitem{Pedram:2012ui}
P.~Pedram,
Int. J. Mod. Phys. D \textbf{22}, 1350004 (2013). 

\bibitem{Curado}
E.~M.~F.~Curado and M.~A.~ Rego-Monteiro, J.~Phys. A {\textbf 34} 3253  (2001).

\bibitem{Bosso:2018syo}
P.~Bosso and S.~Das,
Annals Phys. \textbf{396}, 254 (2018). 


\bibitem{Oh}
C.~H.~Oh and K.~Singh,
J. Phys. A \textbf{27}, 5907 (1994).

\bibitem{Quesne}
C.~Quesne and N.~Vansteenkiste,
J. Phys. A \textbf{28} 7019 (1995). 

\bibitem{Plenio}
S.~P.~Kumar and M.~B.~Plenio,
Nature Commun. \textbf{11}, 3900 (2020).

\bibitem{HGF_thesis}
J.~P.~Hannah,
``Identities for the gamma and hypergeometric functions: an overview from Euler to the present'', Master thesis, University of the Witwatersrand, Johannesburg, South Africa (2013).

\bibitem{HGF_Bailey}
W.~N.~Bailey, ``Generalized Hypergeometric Series'', Cambridge Tracts in Mathematics and Mathematical Physics 32, Cambridge University Press, London (1935).

\bibitem{Bagarello}
F.~Bagarello, E.~M.~Curado and J.~P.~Gazeau, J.~Phys.~A: Math. Theor. \textbf {51}, 155201 (2018).

\bibitem{Kober}
M.~Kober,
Phys. Rev. D \textbf{82}, 085017 (2010).


\bibitem{Dossa:2021tiq}
F.~A.~Dossa,
Phys. Scripta \textbf{96}, 105703 (2021). 

\bibitem{Nou}
K.~Nouicer,
Phys. Lett. B \textbf{646}, 63-71 (2007).

\bibitem{Pedrambis}
P.~Pedram,
Phys. Lett. B \textbf{714}, 317-323 (2012).

\bibitem{Chung}
W.~S.~Chung and H.~Hassanabadi,
Eur. Phys. J. C \textbf{79}, 213 (2019). 

\bibitem{BosTodo}
V.~Todorinov, P.~Bosso and S.~Das,
Annals Phys. \textbf{405}, 92-100 (2019)

\bibitem{Caves}
C.~M.~Caves,
Phys. Rev. D \textbf{23}, 1693-1708 (1981).

\end{thebibliography}
\end{document}